\documentclass[iop,apj,tighten,numberedappendix,appendixfloats,twocolappendix]{emulateapj}

\usepackage{apjfonts}
\usepackage{amsmath}

\newcommand{\rb}[1]{\raisebox{1.1ex}[0pt]{#1}}

\newcommand{\civ}{{\sc C iv}}
\newcommand{\caii}{Ca {\sc ii}}
\newcommand{\feii}{Fe {\sc ii}}

\newcommand{\hb}{H$\beta$}
\newcommand{\hg}{H$\gamma$}
\newcommand{\hd}{H$\delta$}
\newcommand{\hbb}{${\rm H}\beta_{\rm BC}$}
\newcommand{\heii}{He {\sc ii}}
\newcommand{\kms}{$\rm km~s^{-1}$}
\newcommand{\ks}{$\chi^2$}
\newcommand{\neiii}{[Ne {\sc iii}]}
\newcommand{\oii}{{\sc [O ii]}}
\newcommand{\oiii}{[\textsc{O iii}]}
\newcommand{\rfe}{$R_{\rm Fe}$}
\newcommand{\vfe}{$v_{\rm Fe}$}

\newcommand{\myemail}{huc@ihep.ac.cn}

\slugcomment{Accepted for publication in \textit{The Astrophysical Journal}}

\shorttitle{Two-Component \hb. I. SPCA}
\shortauthors{Hu et al.}

\begin{document}

\title{Two-Component Structure of the \hb\ Broad-Line Region in Quasars. I.\\
Evidence from Spectral Principal Component Analysis}

\author{Chen Hu\altaffilmark{1}, Jian-Min Wang\altaffilmark{1,2}, Luis C.
Ho\altaffilmark{3}, Gary J. Ferland\altaffilmark{4}, Jack A.
Baldwin\altaffilmark{5}, and Ye Wang\altaffilmark{4}}

\altaffiltext{1}{Key Laboratory for Particle Astrophysics,
Institute of High Energy Physics, Chinese Academy of Sciences, 19B Yuquan
Road, Beijing 100049, China; \myemail}

\altaffiltext{2}{National Astronomical Observatories of China,
Chinese Academy of Sciences, 20A Datun Road, Beijing 100012, China.}

\altaffiltext{3}{The Observatories of the Carnegie Institution for Science, 
813 Santa Barbara Street, Pasadena, CA 91101, USA}

\altaffiltext{4}{Department of Physics and Astronomy, 177 Chemistry/Physics
Building, University of Kentucky, Lexington, KY 40506, USA}

\altaffiltext{5}{Department of Physics and Astronomy, 3270 Biomedical Physical
Sciences Building, Michigan State University, East Lansing, MI 48824, USA}

\begin{abstract}
  We report on a spectral principal component analysis (SPCA) of a sample of
  816 quasars, selected to have small \feii\ velocity shifts with spectral
  coverage in the rest wavelength range 3500--5500 \AA. The sample is
  explicitly designed to mitigate spurious effects on SPCA induced by \feii\
  velocity shifts. We improve the algorithm of SPCA in the literature and
  introduce a new quantity, \emph{the fractional-contribution spectrum}, that
  effectively identifies the emission features encoded in each eigenspectrum.
  The first eigenspectrum clearly records the power-law continuum and very
  broad Balmer emission lines. Narrow emission lines dominate the second
  eigenspectrum. The third eigenspectrum represents the \feii\ emission and a
  component of the Balmer lines with kinematically similar intermediate
  velocity widths. Correlations between the weights of the eigenspectra and
  parametric measurements of line strength and continuum slope confirm the
  above interpretation for the eigenspectra. Monte Carlo simulations
  demonstrate the validity of our method to recognize cross talk in SPCA and
  firmly rule out a single-component model for broad \hb. We also present the
  results of SPCA for four other samples that contain quasars in bins of
  larger \feii\ velocity shift; similar eigenspectra are obtained. We propose
  that the \hb-emitting region has two kinematically distinct components: one
  with very large velocities whose strength correlates with the continuum
  shape, and another with more modest, intermediate velocities that is closely
  coupled to the gas that gives rise to \feii\ emission.
\end{abstract}


\section{Introduction}

\subsection{\hb\ Broad-line Region}
\label{sec-introhb}
The structure of the broad-line region (BLR) in active galactic nuclei (AGNs)
is still poorly understood. A widely accepted concept, predicted from
photoionization models \citep{collin88} and supported by reverberation mapping
observations \citep[e.g.,][]{peterson99}, is that the BLR is radially
stratified: high-ionization lines are emitted from smaller radii than
low-ionization lines. High-ionization lines such as \civ\ are thought to be
emitted, at least in part, from an outflow \citep[see][and references
therein]{richards11}, while low-ionization lines such as \hb\ originate from a
virialized region.  It is the virialized component that is pertinent to
efforts to use the BLR to estimate the mass of the central black hole (BH).
However, velocity-resolved reverberation data from recent monitoring programs
indicate that \hb-emitting region is more complicated than previously thought;
depending on the object, infall, outflow, and virialized motions are all
possible \citep[e.g.,][and references therein]{bentz10,denney10}.

The profile of the broad \hb\ line also points to the complexity of the
\hb-emitting region. It generally cannot be well described by a single
Gaussian. Two Gaussians \citep[e.g.,][]{netzer07,hu08a} or a Gaussian-Hermite
function \citep[e.g.,][]{salviander07,hu08b} are often used for quasars, while
a Lorentzian, a Lorentzian plus a very broad Gaussian
\citep[e.g.,][]{veron04}, or two Gaussians \citep[e.g.,][]{mullaney08} are
used for narrow-line Seyfert 1 galaxies. In addition, the \hb\ profile shows
great diversity from object to object \citep[e.g.,][and references
therein]{hu08a,zamfir10}. Some previous studies
\citep[e.g.,][]{brotherton96,sulentic00b} propose a two-component model for
\hb\ emission, an intermediate-width component and a very broad component.
Netzer \& Marziani's (2010) calculations of the line profile rule out simple
single-zone models.

\citet{hu08a,hu08b} systematically investigated \feii\ and \hb\ emission in a
large sample of quasars selected from the Sloan Digital Sky Survey (SDSS;
\citealt{york00}). They found that \feii\ emission originates from an
intermediate-velocity region, located farther out from the center,  whose
dynamics may be dominated by infall. The broad \hb\ line can be decomposed
into two physically distinct components, one associated with the conventional
BLR and another with the intermediate-velocity region identified through
\feii. \citet{ferland09} calculated the outward emission from infalling clouds
and found that such a scenario can reproduce the observed \feii\ emission.
These studies demonstrate that fruitful insights can be gained from a
self-consistent investigation of the kinematics of different emission lines.

However, all the above work suffers from a major drawback: the derived
parameters for both the continuum and the emission lines are model-dependent.
The profiles of the emission lines are not necessarily well represented by the
simple analytical functions (Gaussians, Lorentzians, Gauss-Hermite
polynomials, or various combinations thereof) that are commonly used to fit
them. Spectral principal component analysis is an alternative,
model-independent approach that can be used to study emission-line profiles.
This method makes use use of all the emission features in the spectra, and it
is well-suited for application to large data sets, such as that afforded by
SDSS.

\subsection{Spectral Principal Component Analysis}
\label{sec-introspca}

Principal component analysis (PCA) is a powerful mathematical tool used to
reduce the dimensionality of a data set. It describes the variation in the
data set by the fewest number of variables, called principal components (PCs).
The method has been widely used for many purposes in many areas of astronomy,
including studies on AGNs. The most common implementation is to apply it to a
set of measured variables and seek correlations among them.
\citet[][hereinafter BG92]{bg92} applied it to the Palomar-Green sample of
low-redshift quasars and discovered that the bulk of the variance in the
sample is dominated by the inverse correlation between the strengths of the
\feii\ and \oiii\ lines, commonly called Eigenvector 1 (EV1).

Although applying PCA to a set of measured variables is suitable for
multivariate correlation analysis, it has the shortcoming that measuring the
input variables is itself model-dependent. The parameters of emission lines
are often measured from fits using mathematically convenient functions. Some
parameters, such as the shape and asymmetry, can be quantified without fitting
the line, but fitting the continuum and \feii\ emission first is always
necessary. This shortcoming can be mitigated by spectral principal component
analysis (SPCA).

In SPCA, originally developed by \citet{francis92}, instead of using measured
variables, the fluxes in each wavelength bin are used as input variables.
SPCA obviates the need to model the continuum, the pseudocontinuum (due to
blended \feii\ emission), or the line profiles. The resultant PCs are linear
combinations of the fluxes in each wavelength bin, so have the form of spectra
and are represented as eigenspectra hereafter. SPCA has been used very
successfully in classification \citep[e.g.,][]{francis92,yip04b}, in
establishing empirical templates and reconstruction
\citep[e.g.,][]{hao05,boroson10}, and in exploring ``outliers''
\citep[e.g.,][]{boroson10}. It has also been widely adopted to study the
physics of quasars, by means of interpreting the first few eigenspectra. Table
\ref{tab-pcarevsamp} summarizes previous SPCA studies on AGNs, including their
parent sample, number of sources, wavelength range of the eigenspectra, and,
most importantly, their sample selection criteria, which affect the results.
Table \ref{tab-pcarevinter} lists the interpretations of the first three
eigenspectra.

\begin{deluxetable*}{lcccr@{--}ll}
  \tablewidth{0pt}
  \tablecolumns{9}
  \tablecaption{The Samples in Previous SPCA Studies
  \label{tab-pcarevsamp}}
  \tablehead{
  \colhead{} & \colhead{} & \colhead{Parent} &\colhead{Number of} &
  \multicolumn{2}{c}{Wavelength} & \colhead{} 
  \\
  \colhead{\rb{Notation}} & \colhead{\rb{Paper}} & \colhead{sample} &
  \colhead{sources} & \multicolumn{2}{c}{range (\AA)} 
  &\colhead{\rb{Notes on sample selection}} }
  \startdata
  F92 & \citet{francis92} & LBQS & 232 & 1150&2000 & Wavelength coverage\\
  S03A & \citet{shang03} & BQS & 22 & 4000&5500 & Low redshift and Galactic
  absorption \\
  S03B & \citet{shang03} & BQS & 22 & 1171&6608 & Same as above\\ 
  Y04 & \citet{yip04b} & SDSS & 16707 & 900&8000 & 
  All in SDSS DR1 quasar catalog\\
  L09ALL & \citet{ludwig09} & SDSS & 9046 & 4000&6000 & 
  FWHM(\hbb) $> 2000$ \kms \\
  L09S1 & \citet{ludwig09} & SDSS & 6317 & 4000&6000 & Low EW(\oiii) \\
  L09S2 & \citet{ludwig09} & SDSS & 2307 & 4000&6000 & 
  Intermediate EW(\oiii) \\
  L09S3 & \citet{ludwig09} & SDSS & 422 & 4000&6000 & High EW(\oiii) \\
  B10 & \citet{boroson10} & SDSS & 1039 & 4000&5700 &
  Wavelength coverage and spectral quality \\ 
  H12 & This work & SDSS & 816 & 3500&5500 & Small \feii\ velocity shifts
  \enddata
\end{deluxetable*}

\begin{deluxetable*}{llll}
  \tablewidth{0pt}
  \tablecolumns{4}
  \tablecaption{Eigenspectra Interpretations in Previous SPCA Studies
  \label{tab-pcarevinter}}
  \tablehead{
  \colhead{} & \multicolumn{3}{c}{Interpretations of Eigenspectra} 
  \\
  \cline{2-4}
  \colhead{\rb{Notation}}  &\colhead{1st} & \colhead{2nd} & \colhead{3rd} }
  \startdata
  F92 & Emission-line cores & Continuum slope & Broad absorption lines\\
  S03A & BG92's EV1 & Balmer lines & \nodata \\ 
  S03B & Emission-line cores & Continuum slope & Line-width relationships \\ 
  Y04 & Host galaxy component\tablenotemark{a} & 
  Continuum slope\tablenotemark{a} & Balmer emission lines\tablenotemark{a} \\
  L09ALL & Narrow emission lines & Broad emission lines and continuum slope &
  \nodata \\
  L09S1 & Broad emission lines and continuum slope & 
  Correlation among all emission & BG92's EV1 \\
  L09S2 & Broad emission lines and continuum slope & Narrow emission lines &
  Narrow emission-line shift \\
  L09S3 & Narrow emission-line EW & Narrow emission-line shift or asymmetry
  & Narrow emission-line width \\
  B10\tablenotemark{b} & BG92's EV1\tablenotemark{a} & \nodata & \nodata \\
  H12 & Continuum and very broad emission lines & Narrow emission lines &
  Intermediate-width emission lines
  \enddata
  \tablenotetext{a}{\citet{yip04b,boroson10} did not subtract a mean spectrum
  and thus their first eigenspectrum is the mean spectrum. Their $n$th ($n \ge
  2$) eigenspectra are labeled as ($n-1$)th in the present paper.}
  \tablenotetext{b}{\citet{boroson10} did not aim to interpret the physical
  meaning of the eigenspectra, other than equating the first eigenspectrum
  with BG92's EV1.}
\end{deluxetable*}

A disadvantage of using SPCA to explore the physics of AGN emission stems from
the fact that SPCA is a linear analysis, while there are many nonlinear
factors in the distribution of samples and also in the variance of quasar
spectra. These nonlinear factors render very difficult simple physical
interpretation of the eigenspectra. Many eigenspectra presented in the
literature do not show clear and clean features. As listed in Table
\ref{tab-pcarevinter}, some interpretations have ambiguous physical meaning,
and many vary from study to study. 

\vspace{1em}

In the present paper, the first in a series, we aim to derive a set of
eigenspectra of quasar optical spectra that has clear physical meaning, by
considering the nonlinear factors described in \S \ref{sec-pcanonl},
especially concerning the diversity of \feii\ velocity shifts. Our samples are
established in \S \ref{sec-sample}. In \S \ref{sec-pcamath}, we briefly
describe our SPCA algorithm and the definition of a new quantity, the
fractional-contribution spectrum, which helps to understand the eigenspectra;
more details are given in the Appendices. For the sample of quasars with small
\feii\ velocity shift, Section \ref{sec-pcaresult} presents the eigenspectra
we obtain, their physical interpretation, correlations between the weights and
spectral measurements, bootstrap study, and fitting of eigenspectrum 3. We
test our SPCA method and interpretation using Monte Carlo simulations in \S
\ref{sec-sim}. Section \ref{sec-ovbin} presents the results for four other
quasar samples with larger \feii\ velocity shifts. The implications of our
results are discussed in \S \ref{sec-dis}, with a summary given in \S
\ref{sec-sum}.

In the second paper of this series, we will explore the application of
eigenspectrum 3 as a template for \feii\ and for the intermediate-width
component of the Balmer lines. The two-component structure of the \hb\ BLR
suggested by our SPCA results is relevant to BH mass estimates using the \hb\
line width. This issue will be investigated in a third paper.

\section{Sample Selection}
\label{sec-samsel}

We use the sample defined in \citet{hu08b} as the parent sample. It is the
largest sample to date that has measurements of \feii\ velocity shifts. The
parent sample was selected from the SDSS Fifth Data Release (DR5;
\citealt{adelman07}) quasar catalog \citep{schneider07} according to a series
of criteria that ensure that reliable measurements can be obtained for \feii\
emission. Briefly, the selection criteria include: (1) redshift $z < 0.8$ and
signal-to-noise ratio (S/N) $> 10$ in the wavelength range 4430--5550 \AA; (2)
\ks\ $< 4$ in their continuum decomposition; (3) equivalent width (EW) of
\feii\ $> 25$ \AA; (4) the \hb\ broad component FWHM errors $< 10$\% and
\oiii\ $\lambda$5007 peak velocity shift errors $<$100 \kms. 

\subsection{Nonlinear Factors in SPCA}
\label{sec-pcanonl}

Many nonlinear factors in SPCA have been studied in the literature. Variable
line width is one of them. The line center and wing have opposite responses
when varying only the line width. Its influence on SPCA has been studied
extensively using simulations \citep[e.g.,][]{mittaz90,shang03}. It produces
the characteristic Fourier-like ``W'' shape on eigenspectra, a feature that
has been seen in the majority of SPCA studies on quasar spectra. Another
nonlinear factor that contributes to the variance of quasar spectra is the
variable slope of the power-law continuum. \citet{shang03} studied the
influence of the continuum slope and found that it introduces cross talk among
eigenspectra, which manifests as the appearance of a spectral feature (e.g.,
an emission line) on an eigenspectrum with which it does not correlate.

The nonlinear effects caused by these variable parameters can be mitigated by
judiciously selecting subsamples in which these parameters (e.g., \hb\ line
width) span a narrow range of values.  However, as discussed in \S
\ref{sec-introhb}, the \hb\ profile is too complicated to be described by a
single width parameter.  Moreover, one of our primary goals is to study the
\hb\ profile itself, and it would be counterproductive to restrict our sample
using the very parameter we wish to investigate. We could attempt to subtract
the power-law continuum first, but this, too, is not ideal because the
continuum cannot be measured in a model-independent manner, thereby
compromising the advantages to be gained by SPCA.

The spectra of quasars also vary in the velocity shifts of emission lines.
Broad quasar emission lines often show considerable velocity shifts with
respect to narrow emission lines \citep{gaskell82}. The velocity shift varies
from source to source and from line to line (see, e.g., introduction section
of \citealt{hu08b} for a brief review). \citet{hu08b} systematically
investigated the optical \feii\ emission in a large sample of quasars selected
from SDSS, and found that the velocity shifts of \feii\ emission span over a
wide range. The majority of quasars show \feii\ emission that is redshifted by
$\sim$ 0 to 1000 \kms, some up to 2000 \kms, with respect to the velocity of
the narrow-line region as traced by \oiii\ $\lambda$5007 or \oii\
$\lambda$3727 (see Figure 7 of \citealt{hu08b}). 

Recently, \citet{sulentic12} called into question the measurement of \feii\
velocity shifts in \citet{hu08b} by testing the composite spectra of some
subsamples. They claimed that the \feii\ emission in their composite spectra
do not show significant%
\footnote{
One composite spectrum (B1) in \citet{sulentic12} has a best-fit value of
\feii\ velocity shift of 730 \kms. They claimed that the shift is not
distinguishable from zero by adopting the criterion that $\chi^2/\chi^2_{\rm
min}$ should be larger than 1.24. Their criterion is much too severe than that
derived from $F$-test of an additional term (free versus fixed \feii\ velocity
shift) in the fitting. See Appendix \ref{app-feschi} for details.
}
velocity shifts, and concluded that there is no evidence for a systematic
\feii\ redshift. They preferred the older conclusion that the velocity shift
and width of \feii\ follow those of \hb\ in virtually all sources. However,
the properties of a composite spectrum depend on how the subsample is
selected. Contrary to the composite spectra in \citet{sulentic12} defined by
their 4D Eigenvector 1 formalism, the composite spectra of quasars in bins of
different \feii\ velocity shifts generated in \citet{hu08b} \emph{do} show
significantly redshifted \feii\ emission. Appendix \ref{app-feschi} gives
details. The results from composite spectra in Appendix \ref{app-feschi}
provide additional evidence that the \feii\ velocity shift measurements in
\citet{hu08b} are reliable for the majority of sources.

The velocity shift of the emission line is a promising nonlinear factor, but
one that has been overlooked in previous SPCA studies. Prior to any analysis,
all the spectra have been deredshifted to their rest frame. If the sources
have different intrinsic velocity shifts, the effect on SPCA is equivalent to
a situation in which the sources have no velocity shifts but their spectra are
deredshifted using the \emph{wrong} redshifts. Both cases will confuse the
resultant eigenspectra, rendering any simple interpretation impossible. This
effect can be avoided if we restrict the variance of the velocity shifts in
the sample to be small---much smaller than the widths of the broad emission
lines. As this condition is satisfied by the narrow lines and broad Balmer
lines in the optical spectra of quasars, it is reasonable that previous SPCA
studies did not take velocity shifts into consideration. \feii\ emission,
which had not been well studied, was simply assumed to have no large velocity
shift. But, as mentioned above, the dispersion in \feii\ emission shifts is
comparable to the full width at half-maximum (FWHM) of the line (median value
$\sim$ 2500 \kms; Figure 8 of \citealt{hu08b}). Thus, the velocity shift of
optical \feii\ emission \emph{cannot} be ignored in SPCA.

To summarize this subsection: the variance in \feii\ emission velocity shifts
in quasars is an important nonlinear factor. It must be treated carefully in
any SPCA study. To minimize the adverse effects of this nonlinear factor, we
will construct subsamples in which \feii\ emission has similar velocity
shifts.

\subsection{Our SPCA Sample}
\label{sec-sample}

We select the primary sample for SPCA from the parent sample using the
following additional criteria. (1) We choose objects with $z < 0.67$ in order
to obtain eigenspectra covering the rest-frame wavelength range 3500--5500
\AA. The red end ensures that we include the \feii\ emission redward of \oiii\
$\lambda$5007 in the eigenspectra. (2) The velocity shifts of \feii\ emission
are in the range $-$250 to 250 \kms, to ensure that the influence of this
nonlinear factor can be suppressed. (3) Following \citet{boroson10}, we select
spectra that have no more than 100 pixels flagged as bad by the SDSS pipeline
in the rest-frame wavelength range 3500--5500 \AA. The final primary sample
for SPCA contains 816 sources, all of which have small \feii\ velocity shifts.

The \feii\ velocity shift limits adopted in criterion (2) above follows that
for subsample A in \S 4.6 of \citet{hu08b}. Four other bins of \feii\ velocity
shift were use in \citet{hu08b} (see their Figure 12). In this paper, we also
establish four other subsamples for SPCA, to explore the consequences of
varying the velocity shift criterion, while keeping the other criteria
unchanged.  The number of sources in the other four subsamples are 794 (B,
250--750 \kms), 338 (C, 750--1250 \kms), 203 (D, 1250--1750 \kms), and 110 (E,
1750--2250 \kms).

The results of SPCA for the four bins with large \feii\ velocity shifts are
presented only in \S \ref{sec-ovbin}. Unless otherwise noted, the SPCA results
in this paper refers to the primary sample of 816 quasars with \feii\ velocity
shift in the range $-$250 to 250 \kms\ (sample A).

\section{Methodology}
\label{sec-pcamath}

Except for employing the same algorithm to obtain the eigenspectra from the
cross-correlation matrix, previous studies were different in many aspects.
Table \ref{tab-pcarevsamp} lists the main differences among previous studies,
including sample selection and wavelength coverage. These are the main reasons
that they obtained different eigenspectra and interpretations.  Additional
differences arise from different implementation of a number of technical
details in the analysis, such as whether the mean spectrum is subtracted or
not, how the spectra are normalized, and how to deal with noise and gaps in
the data.

The present work incorporates two major improvements. (1) We isolate a sample
of quasars with a narrow range of \feii\ velocity shifts to minimize nonlinear
effects in SPCA. (2) We introduce a new quantity, called the
fractional-contribution spectrum; this parameter will turn out to be useful in
understanding the physical meaning of the eigenspectra.  Our technique adopts
a slightly different method of normalizing the spectra compared to previous
studies.

We apply SPCA to the 816 quasar spectra selected above. The process of
constructing the cross-correlation matrix mainly follows \citet{francis92}.
In detail, the steps are as follows.

\vspace{1em}

1. Each spectrum is corrected for Galactic extinction using the
\textit{u}-band extinction listed in \citet{schneider07} and the extinction
law of \citet{cardelli89} and \citet{odonnell94}. We then shift the spectrum
to its rest frame using the redshift given in \citet{hu08b}, which is based on
\oiii\ $\lambda$5007. Although \oiii\ can be blueshifted, we adopt it to
define the zero point of the velocity because the measurements of \feii\
velocity shift were calculated with respect to \oiii\ in \citet{hu08b} (see
their \S 3.5 for details). After deredshifting, all the spectra are rebinned
to fixed wavelength bins, which are equally divided in logarithmic space with
a dispersion of $d\lambda/\lambda = 10^{-4}$, covering the range 3500--5500
\AA. The dispersion is selected to be equal to that of SDSS spectra, and the
wavelength range is covered by all the spectra in our sample. There are 1962
wavelength bins for each rebinned spectrum.

2. The rebinned spectrum is normalized to unity average flux density over the
wavelength interval 5075--5125 \AA. Normalization by the scalar product of the
spectrum \citep[e.g.,][]{yip04b,boroson10} or by the integrated flux
\citep[e.g.,][]{francis92,shang03,ludwig09} were adopted in previous work. We
tested these three alternative methods of normalization and found that the
derived eigenspectra are similar (see also the test in \S 3.2 of
\citealt{connolly95}). However, we found that the normalization method adopted
here (to have the unity flux density at 5100 \AA) produces more meaningful
fractional-contribution spectrum (see Appendix \ref{app-pov} for the
definition), which helps us to interpret the physical meaning of the
eigenspectra. Following \citet[][see their \S 3.2]{francis92}, we do not
scale the flux values for the sample in each wavelength bin to unity variance.

3. We calculate the mean spectrum of the normalized spectra using all the good
pixels and then subtract it from each of the original spectra. Some previous
studies do not subtract the mean spectrum \citep[e.g.,][]{yip04b,boroson10}.
In that case, the first eigenspectrum is equal to the mean spectrum, and their
$n$th ($n\ge2$) eigenspectrum should be compared with the ($n-1$)th
eigenspectrum in the present paper (see also Table \ref{tab-pcarevinter}).

\vspace{1em}

The above steps result in a mean-subtracted flux matrix $\mathbf{F}$, whose
element $f_{ij}$ is the flux of the $i$th mean-subtracted spectrum in the
$j$th wavelength bin. The goal of SPCA is to calculate the eigenvalues and
eigenvectors of the cross-correlation matrix $\mathbf{F}^T \cdot \mathbf{F}$,
where $\mathbf{F}^T$ denotes the transpose of matrix $\mathbf{F}$. This is
achieved using singular value decomposition. The iteration method of
\citet{yip04a} is adopted to deal with the noise and bad pixels in the
spectra. Then, the mean-subtracted spectra are reconstructed as $\mathbf{F} =
\mathbf{W} \cdot \mathbf{V}^T$, where $\mathbf{V}$ has columns $\mathbf{v}_k$
as the eigenspectra, $\mathbf{W}$ is an matrix whose element $w_{ik}$ is the
weight of the $k$th eigenspectrum for the $i$th mean-subtracted spectrum. The
algorithm is described in detail in Appendix \ref{app-derivation}.

The eigenspectra are arranged in the order of decreasing eigenvalues
$\lambda_k^2$, which are also used to calculate the contribution of each
eigenspectrum to the total variation of the input spectra as $P_k =
\lambda_k^2 / \sum_l \lambda_l^2$ \citep{mittaz90}. Thus, the first $n$th
eigenspectra can account for $\sum_{k=1}^n P_k$ percentage of the total
variation. This quantity is used to determine how many eigenspectra are
sufficient for understanding the input spectra, by comparing it to the
intrinsic variation in percentage $P^{\rm int}$ \citep{francis92}. The
quantities above are widely used in previous SPCA studies, but only have
meaning as an average over the entire wavelength range. In this work, we
develop them into more sophisticated forms that apply to each wavelength bin,
which are helpful in interpreting the eigenspectra in a physical way. 

We calculate the intrinsic variation and residual variation in each wavelength
bin as \citet{francis92} did, but do not integrate them over wavelength. In an
arbitrary $j$th wavelength bin, we define $p^{\rm int}_j$ as the proportion of
the intrinsic variation, and $p_{jk}$ represents the proportion of the
variation accounted for by the $k$th eigenspectrum in this particular
wavelength bin. Thus, we obtain a matrix $\mathbf{P}$, which has the same
shape as $\mathbf{V}$. Each column ($\mathbf{p}_k$) of $\mathbf{P}$ represents
the contribution of the corresponding eigenspectrum and has the form of a
spectrum.  We call it \emph{the fractional-contribution spectrum}. Using the
fractional-contribution spectrum, it is straightforward to understand which
spectral feature is dominated by which eigenspectrum. The details of our
definitions and the comparison to previously used quantities are presented in
Appendix \ref{app-pov}.

\section{Results}
\label{sec-pcaresult}

In this section, we will first describe and interpret the derived eigenspectra
phenomenologically, and then we will investigate correlations between the
weights of the eigenspectra and the actual measurements derived from the
original spectra, to confirm the interpretation. Bootstrap is performed to
study the stability and uncertainty of the eigenspectra. Lastly, we will fit
eigenspectrum 3 to demonstrate that it consists of a single,
intermediate-width emission-line component.

\subsection{Eigenspectra: Description}
\label{sec-pcades}

\begin{figure}
  \centering
  \includegraphics[width=0.45\textwidth]{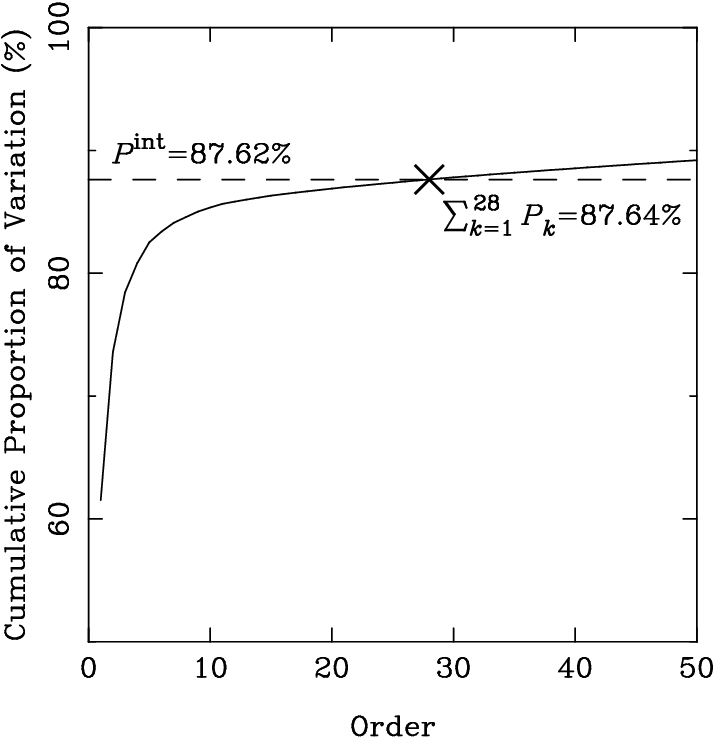}
  \caption{Cumulative proportion of variation accounted for by eigenspectra of
  the first $n$th order. The horizontal dashed line marks the intrinsic
  variation in percentage, and the cross marks where the cumulative proportion
  of variation exceeds the intrinsic variation.}
  \label{fig-cumpov}
\end{figure}

Figure \ref{fig-cumpov} shows the cumulative proportion of variation accounted
for by eigenspectra added successively. The intrinsic variation, in
percentage, is $P^{\rm int}=87.62\%$ (horizontal dashed line); it can be
accounted for by the first 28 eigenspectra ($\sum_{j=1}^{28}P_j=87.64\%$).
This confirms that it is sufficient to considering only the first 30
eigenspectra (Appendix \ref{app-derivation}).

\begin{figure*}
  \centering
  \includegraphics[width=0.82\textwidth]{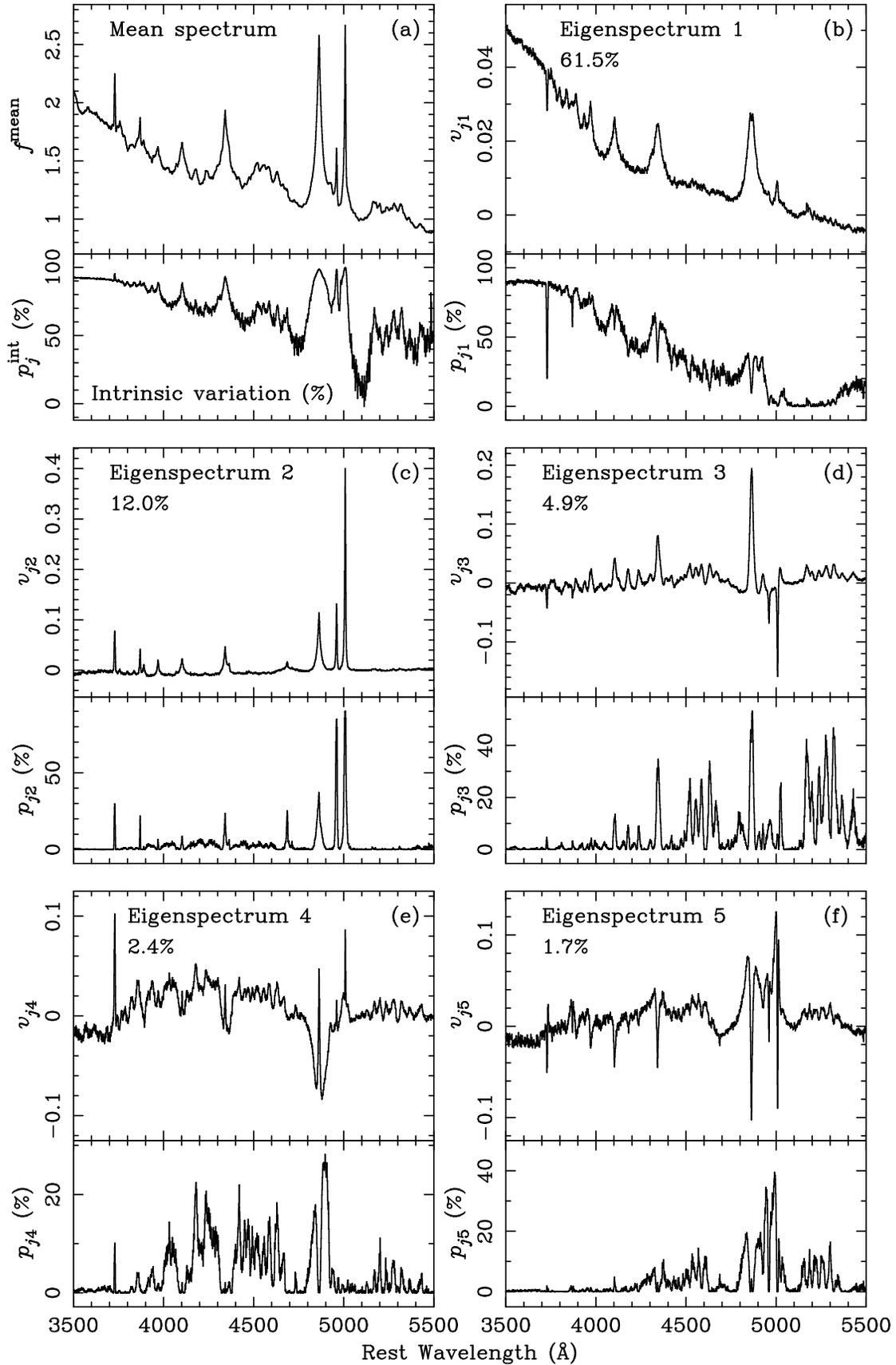}
  \caption{(a) Mean spectrum (upper panel) and proportion of the intrinsic
  variation (lower panel). (b)--(f) First five eigenspectra (upper panels) and
  the corresponding fractional-contribution spectra (lower panels). The
  numbers in the upper panels indicate the contribution of the eigenspectra to
  the total variation over the entire wavelength range (defined in Equation
  [\ref{eqn-pk}]), which is used traditionally.}
  \label{fig-esall}
\end{figure*}

The upper panel of Figure \ref{fig-esall}(a) shows the mean spectrum of the
normalized spectra. As it resembles the composite quasar spectrum of
\citet{vandenberk01}, it implies that the sample for SPCA used here is not
strongly biased.  The lower panel shows, $p_j^{\rm int}$, the proportion of
the intrinsic variation in each wavelength bin.  It drops to 0 around 5100
\AA\ because we normalized the spectrum to unity over 5075--5125 \AA.
$p_j^{\rm int}$ rises toward shorter wavelengths because of the blue power-law
continuum of quasars.  As expected, $p_j^{\rm int}$ also shows that the most
variable emission features are the Balmer lines, \oiii, and \feii.

Figures \ref{fig-esall}(b)--(f) show the first five eigenspectra (upper
panels) and the corresponding fractional-contribution spectra (lower panels).
The number in the upper panel indicates the proportion of the variation
accounted for by each eigenspectrum, as defined in Equation (\ref{eqn-pk}).
Apart from the first five shown here, higher order eigenspectra each
contribute only a tiny proportion of the variation ($P_k < 1\%$ for $k \ge
6$). The first three eigenspectra are the most prominent, not only because
they contribute, respectively, 61.5\%, 12.0\%, and 4.9\% of the variation in
total, but also because they account for more than $\sim$50\% of the variation
in many wavelength bins. Moreover, the three eigenspectra resemble realistic
components in quasar spectra.  We will describe them phenomenologically below.

Eigenspectrum 1 clearly consists of two components: a power-law continuum and
Balmer emission lines. The Balmer continuum is also present, because the blue
end of the eigenspectrum is steeper than a simple power law extrapolated from
the red end. The Balmer emission lines in this eigenspectrum are broader than
those in the mean spectrum. The fractional contribution decreases from
$\sim$100\% at the blue end to $\sim$0\% at the red end; considering our
adopted normalization, this indicates that this eigenspectrum represents the
change of the slope of the power-law continuum. In detail, the
fractional-contribution spectrum dips near the center of the Balmer lines,
indicating eigenspectrum 1 contributes only to the line wings.

Eigenspectrum 2 and its fractional-contribution spectrum clearly reveals a
series of narrow emission-line features, including \oii\ $\lambda$3727,
\neiii\ $\lambda$3869, \oiii\ $\lambda$4363, \heii\ $\lambda$4686, \oiii\
$\lambda\lambda$4959, 5007, and Balmer lines.  Note that narrow \heii\ is weak
in the eigenspectrum but prominent in the fractional-contribution spectrum.
This suggests that weak \heii\ also varies and correlates with other narrow
emission lines. Eigenspectrum 2 accounts for almost 100\% at the center of
\oiii\ $\lambda\lambda$4959, 5007, meaning that it causes almost all the
variation in narrow emission lines.

Eigenspectrum 3 accounts for only 4.9\% of the total variation but contributes
significantly in several narrow wavelength bands. Figure \ref{fig-esall}(d)
exhibits clean \feii\ emission, as well as Balmer lines that are narrower than
those in eigenspectrum 1 but broader than those in eigenspectrum 2. The
fractional-contribution spectrum in the lower panel indicates that this
eigenspectrum mainly contributes \feii\ emission and the core of the Balmer
lines. Note that although the eigenspectrum shows negative narrow features at
the wavelengths of \oii, \neiii, and \oiii, it accounts for almost no
variation in these wavelength bins. Thus, these negative features are neither
absorption lines nor inverse correlations between the narrow emission lines
and \feii; they arise only from cross talk \citep{shang03}. Otherwise, as
demonstrated by the simulations in \S \ref{sec-simct}, the fractional
contribution will be large at the wavelengths of these lines.

Eigenspectra 4 and 5 do not resemble realistic spectra, but show ``W''-shaped
features that represent variations in emission-line widths \citep{mittaz90}.
They contributes variation in the same wavelength bands where they show the
``W'' shapes. Eigenspectrum 4 mainly contributes to the variation of \hb\
wings and \feii, indicating the correlation between \hb\ width and \feii\
strength.  Eigenspectrum 5 contributes $\sim$40\% variation at the blue wings
of \oiii\ $\lambda\lambda$4959, 5007, suggesting that it represents the
blueshifted wings of the \oiii\ lines. It also accounts for a fraction of the
\feii\ variation, indicating a correlation between \feii\ emission and the
\oiii\ blue wing.

\begin{figure}
  \centering
  \includegraphics[width=0.45\textwidth]{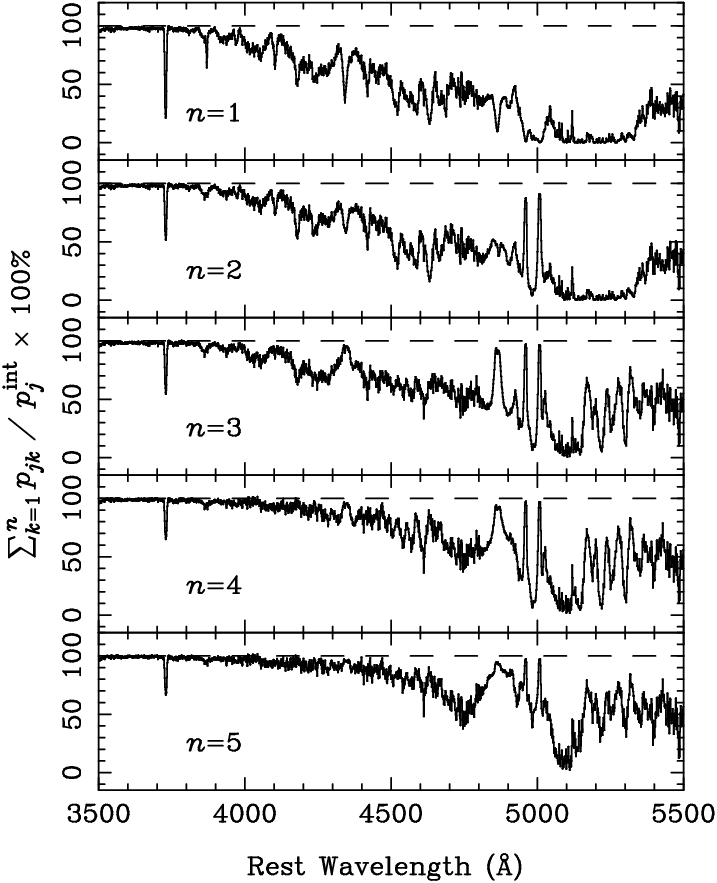}
  \caption{Normalized cumulative fractional-contribution spectra for the first
  five eigenspectra, increasing monotonically from top to bottom. The
  horizontal dashed line in each panel marks 100\%.}
  \label{fig-cfcs}
\end{figure}

Figure \ref{fig-cfcs} shows cumulative fractional-contribution spectra
normalized to the proportion of the intrinsic variation, $\sum_{k=1}^n p_{jk}
/ p_j^{\rm int}$, for the first five eigenspectra. $p_j^{\rm int}$ has been
smoothed with a 3-pixel boxcar filter, to avoid abrupt spikes around 5100 \AA\
in the normalized cumulative fractional-contribution spectra. They reinforce
our interpretation above. Adding eigenspectrum 2 increases the contribution to
narrow emission lines, especially \oiii, from a few to almost 100 percent.
$\sum_{k=1}^3 p_{jk}$, compared to $\sum_{k=1}^2 p_{jk}$, shows significant
\feii\ and intermediate-width Balmer emission lines. Adding eigenspectrum 4
and 5 slightly changes the wings of \hb\ and \oiii, respectively. The
normalized $\sum_{k=1}^n p_{jk}$ already approximate 100\% (the horizontal
dashed line) in many wavelength bins when $n=5$, and reach almost unity in the
entire wavelength range when the first 28 eigenspectra were added (see also
Figure \ref{fig-cumpov}).

In summary, from a phenomenological examination of the eigenspectra and the
corresponding fractional-contribution spectra, the first three eigenspectra
represent the power-law continuum $+$ very broad emission lines, narrow
emission lines, and intermediate-width emission lines, respectively. In the
next subsection, we will investigate the correlations between the weights of
the eigenspectra and some measured variables, to confirm the interpretation of
the eigenspectra above.

\subsection{Eigenspectra: Correlations}

The weights of the eigenspectra used in the reconstruction ($\mathbf{W}$ in
Equation [\ref{eqn-recon}]) have two usages: (1) they can reveal
subpopulations according to their distributions and then be used for
classification \citep[e.g.,][]{francis92,ludwig09,boroson10}; (2) they can be
used to seek correlations with other measured quantities to elucidate their
physical meaning. We follow the latter approach in this paper. 

Among the large number of quantities measured by \citet{hu08b} for their
quasar sample, we explore the power-law spectral index of the continuum
($\alpha$, defined as $f_\lambda \propto \lambda^\alpha$), EW of broad \hb\
[EW(\hbb)], EW of \oiii\ $\lambda$5007 [EW(\oiii)], and EW of the \feii\
emission between 4434 \AA\ and 4686 \AA\ [EW(\feii)]. We choose these three
emission lines because they are the typical lines with broad, intermediate,
and narrow line widths in the optical band \citep{hu08b}. For simplicity, all
EW measurements refer to the continuum at 5100 \AA\ \citep[e.g.,][]{bg92};
thus, they are not exactly the luminosity ratios of the emission lines to the
ionizing continuum (see \S \ref{sec-ewalpha} below for more discussion).
EW(\hbb) is derived from their Gauss-Hermite fitting for the broad \hb\
component, and EW(\oiii) comes from the sum of two Gaussian components if a
second Gaussian is needed. For the sources in our sample, the broad \hb\
component and \oiii\ are fitted very well by a Gauss-Hermite and
double-Gaussian functions, respectively; EW(\hbb) and EW(\oiii) used here
refer to the EWs for the entire emission line and depend little on the model
used for the profile fitting.

\begin{figure*}
  \centering
  \includegraphics[width=0.9\textwidth]{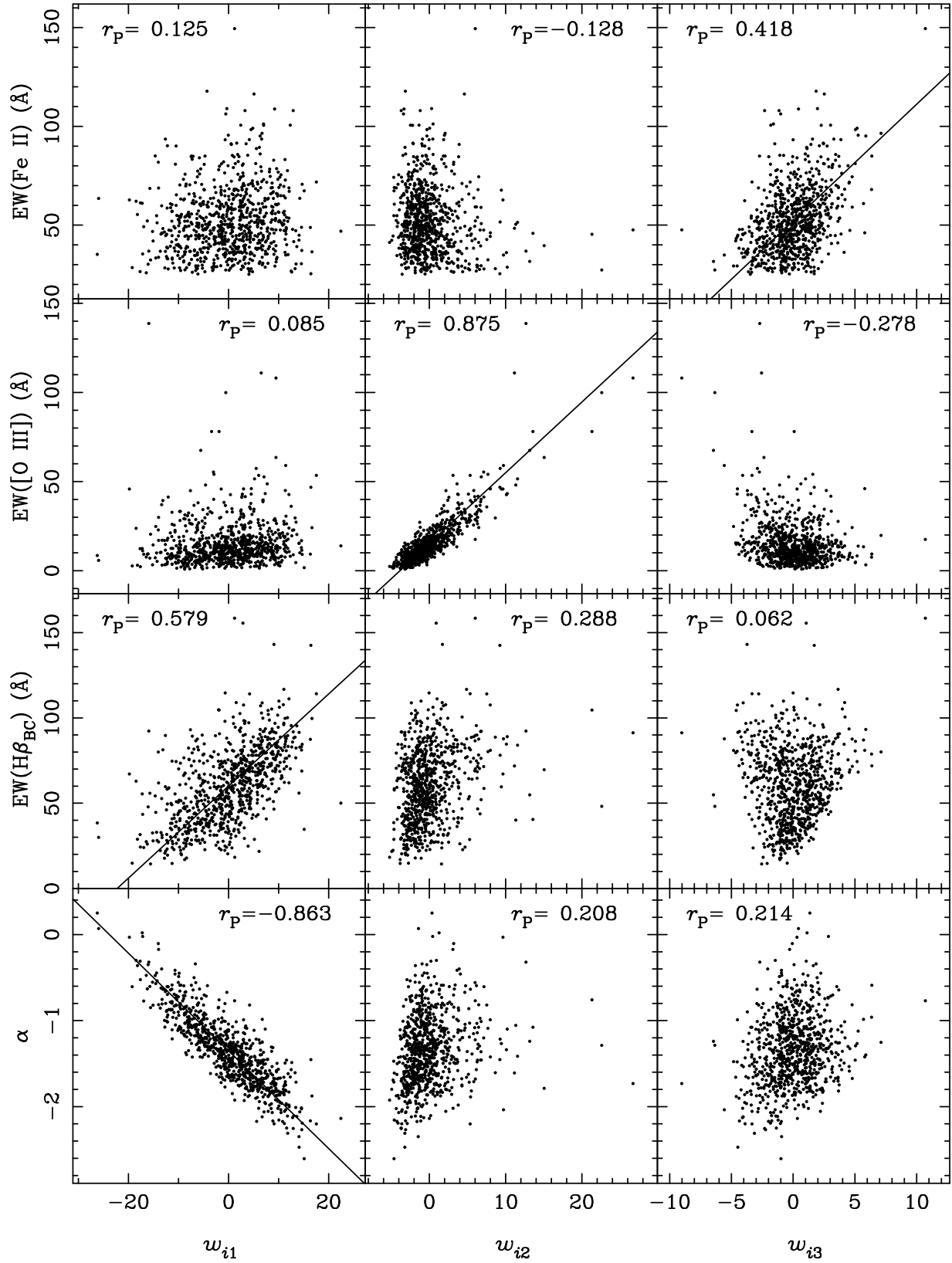}
  \caption{Weights of the first three eigenspectra versus spectral index
  $\alpha$, EWs of the broad component of \hb, \feii, and \oiii. The number in
  each panel is the Pearson's correlation coefficient $r_{\rm P}$ for the two
  quantities plotted. For cases with $r_{\rm P} > 0.4$, the solid line shows
  the fit using OLS bisector method \citep{isobe90}. See text for the
  deviation between the fit and the data points in the top-right panel.}
  \label{fig-corall}
\end{figure*}

Figure \ref{fig-corall} shows the weights of the first three eigenspectra
plotted against the measurements mentioned above. The correlation between each
pair of quantities is tested by calculating the Pearson's correlation
coefficient $r_{\rm P}$, which is given in each panel. The weight of
eigenspectrum 1 strongly correlates with $\alpha$ and EW(\hbb), with $r_{\rm
P} = -0.863$ and 0.579, respectively, supporting the proposition that
eigenspectrum 1 represents both the continuum and the Balmer emission lines.
For the weight of eigenspectrum 2, the strongest correlation is with
EW(\oiii), yielding $r_{\rm P} = 0.875$, while eigenspectrum 3 correlates best
with EW(\feii), with $r_{\rm P} = 0.418$. These results indicate that
eigenspectrum 2 represents narrow emission lines and eigenspectrum 3
corresponds to \feii\ emission. At the same time, that each of the four
measurements strongly correlates with only one of the weights demonstrates the
validity of using these eigenspectra to decompose quasar spectra. The four
strongest correlations are fitted by the OLS bisector method \citep{isobe90},
and the best-fitting fits are shown as solid lines in the corresponding
panels. The four fitted lines match the trends of the data points well. The
only exception is the correlation between $w_{i3}$ and EW(\feii) (top-right
panel), which is somewhat flatter.  This deviation is induced by the bias in
our sample.  Our sample selection excludes objects with EW(\feii) $< 25$ \AA;
this cutoff imposes an artificial sharp angle, which flattens the trend.

The results of the correlation analysis are consistent with the
phenomenological analysis on the eigenspectra and fractional-contribution
spectra presented in the last subsection.

\subsection{Bootstrap: Stability and Uncertainty of the Eigenspectra}
\label{sec-bs}

Outliers with rare features in their spectra (e.g., those with low-ionization
broad absorption lines, which may exist in our sample) affect the results of
SPCA. The instability thus induced in the eigenspectra can be tested using a
bootstrap method \citep[e.g.,][]{wang11}, which is adapted here.  

\begin{figure}
  \centering
  \includegraphics[width=0.43\textwidth]{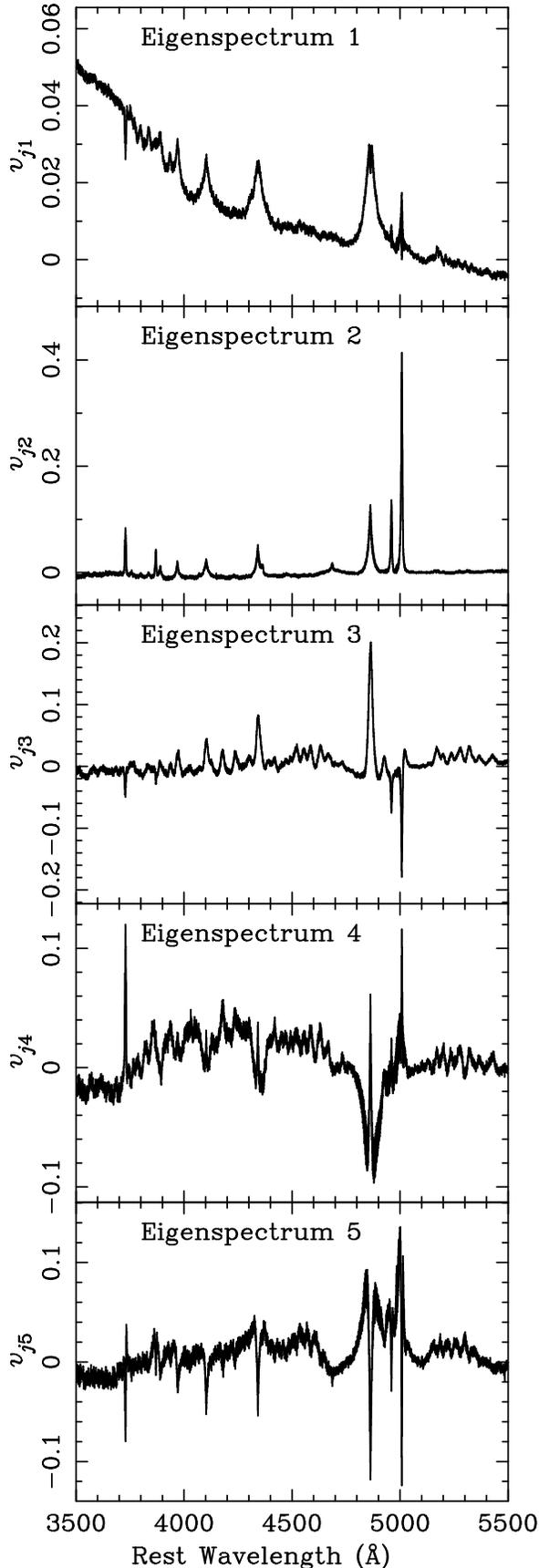}
  \caption{The shaded regions in the panels show the 1$\sigma$ distributions
  of the 100 bootstrap realizations of the original samples for the first 5
  eigenspectra.}
  \label{fig-esbs}
\end{figure}

In each bootstrap realization, a resample of 816 spectra is obtained by random
sampling with replacement from the original sample, such that some spectra
might be absent while some might be present more than once. Then, the same
steps described in \S \ref{sec-pcamath} and Appendix \ref{app-derivation} are
performed to derive the eigenspectra. We repeat the bootstrap 100 times and
obtained 100 series of eigenspectra. Figure \ref{fig-esbs} shows the one
standard deviation of the first five eigenspectra. The eigenspectra are quite
stable, demonstrating that our sample selection is suitable for SPCA. Among
the five, eigenspectra 2 and 3 have the smallest uncertainty, which is
understandable if the two each represent a single emission-line component
(narrow and intermediate-width emission lines, respectively), as we suggest.

The standard deviations of the eigenspectra given by the bootstrap method also
serve as an error estimate, which is helpful for the fitting below.

\subsection{Fitting the Eigenspectra}

As shown in Figure \ref{fig-esall}, for either eigenspectrum 1 or 2, it is
easy to identify the emission lines and conclude that within each
eigenspectrum the lines have roughly the same profile. For eigenspectrum 3,
the Balmer lines are clear and obviously narrower than those in eigenspectrum
1. But while the blended \feii\ lines are significant, the profile of each
single \feii\ line cannot be resolved without fitting. It is not obvious
whether all the \feii\ lines have the same profile, and if so, whether they
have the same profile as the Balmer lines in eigenspectrum 3. Thus, we fit the
mean eigenspectrum 3 given by the bootstrap method in \S \ref{sec-bs}, using
the standard deviation derived by bootstrap as the error, to explore these
questions.

\begin{figure}
  \centering
  \includegraphics[width=0.45\textwidth]{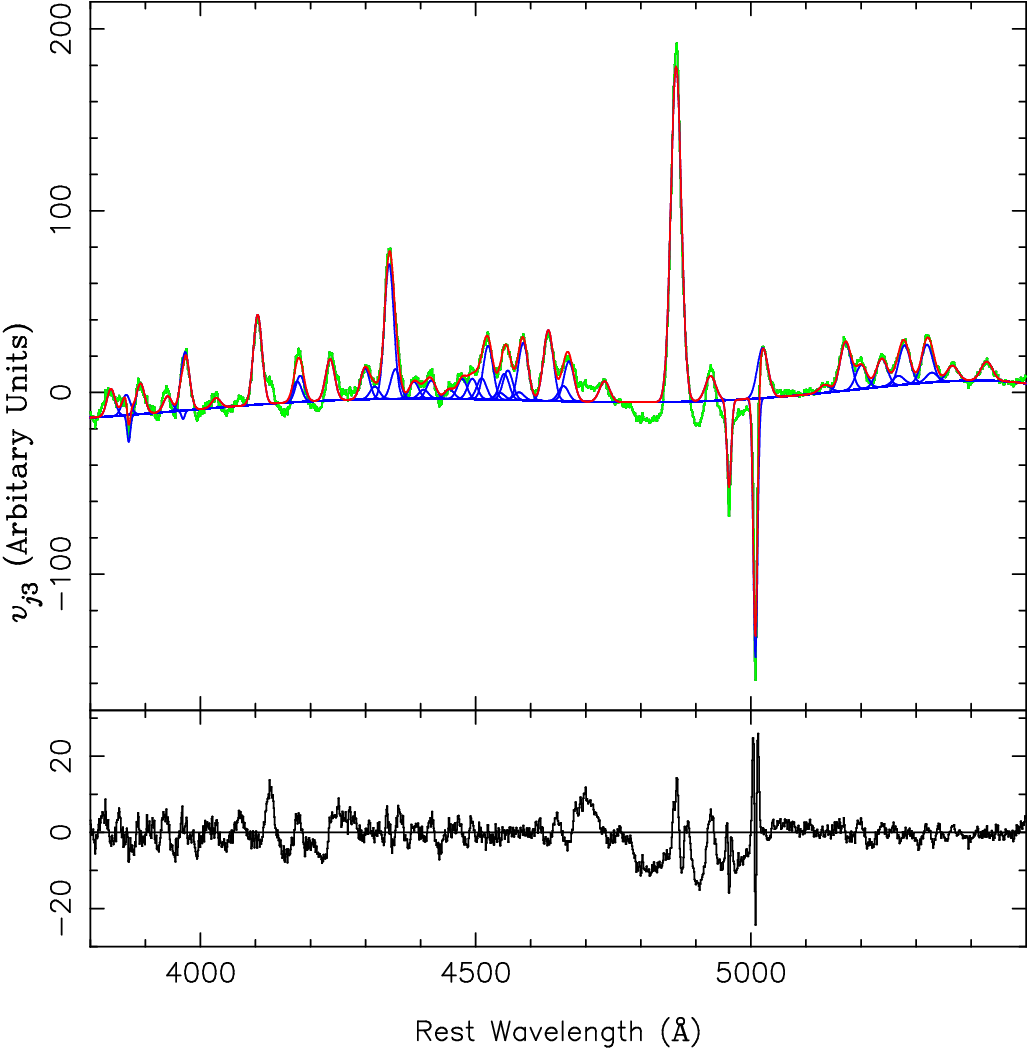}
  \caption{Fitting of eigenspectrum 3. The top panel shows the mean
  eigenspectrum 3 (green) and the model (red), which consists of three
  components (blue), including a Legendre polynomial ``continuum,'' a set of
  Gaussians to represent \feii\ and Balmer lines, and a set of Gaussians with
  negative intensity to represent the \oiii\ lines that arise from cross talk.
  The bottom panel shows the residuals.}
  \label{fig-es3fit}
\end{figure}

As shown in Figure \ref{fig-es3fit}, the mean eigenspectrum 3 can be fitted
well with the following three components. (1) A Legendre polynomial to model
the ``continuum.'' (2) A set of single Gaussians with the same velocity shift
and width, the intensities of which are free to vary. Each Gaussian is an
emission line, including Balmer lines from \hb\ to H$\eta$, and \feii\ lines.
We assemble the line list for \feii\ from \citet{veron04} and \citet{sigut03}.
(3) Another set of lines, each of which is modeled by a single Gaussian with
negative intensity and fixed width and shift, to represent the cross-talk
features at the location of \oiii\ $\lambda\lambda$4959, 5007 and \neiii\
$\lambda$3869. 

Excluding the cross talk component (the third component above), the emission
lines in eigenspectrum 3 can be fitted well using a \emph{single} Gaussian
with identical width and shift.  This finding strongly supports our
proposition, based on the phenomenological analysis in \S \ref{sec-pcades},
that eigenspectrum 3 represents an independent, physically distinct emission
component characterized by intermediate velocities. 

Our fits yield a list of emission-line intensity ratios, one that, as in
\citet{veron04} and \citet{dong11}, can be easily used as a template to model
real quasar spectra. If only the \feii\ lines in the list are used, it works
as an optical \feii\ template. This application will be discussed in the
second paper of this series. A more interesting application is to combine both
the \feii\ and the Balmer lines into a single template for the
intermediate-line region. Such an approach would assume that the \hb\ to
\feii\ intensity ratio is roughly constant in the intermediate-line region for
different sources. This is possible if the clouds that emit
intermediate-velocity lines have large column densities, as suggested by
\citet{ferland09}, who predict that the outward emission reaches an asymptotic
value of \feii/\hb\ $\approx$ 3 (see their Figure 3). If this assumption is
tenable, the \feii\ emission can be used to constrain the intermediate-width
component of \hb, thereby enabling a more physically motivated strategy for
decomposing the \hb\ profile. This is relevant for black hole mass estimation
and will be explored in the third paper of this series.

\section{Simulations}
\label{sec-sim}

Simulations with artificial spectra are used in the literature to test the
validity and limitations of using the SPCA method for physical interpretation
\citep[e.g.,][]{mittaz90,brotherton94,shang03}. While SPCA is successful in
recognizing independent sets of correlated emission features, it is widely
acknowledged that the interpretation is severely complicated by cross talk
(e.g., the negative narrow features in our eigenspectrum 3; Figure
\ref{fig-esall}(d)).  More seriously, different eigenspectra, such as SPC1 and
SPC3 in \citet{shang03} or the intermediate-line region component in
\citet{brotherton94}, cannot be uniquely attributed to distinct physical
components. While these eigenspectra can be ascribed to independently varying
components, as we do in our study, this interpretation is not unique. They can
also be produced by a single emission-line component whose width becomes
broader with lower peak flux (see Figure 13 of \citealt{shang03}). 

In \S \ref{sec-pcamath} above, we introduce the fractional-contribution
spectra matrix $\mathbf{P}$ (calculated by Equation [\ref{eqn-pov}] in
Appendix \ref{app-pov}), which, as shown in \S \ref{sec-pcades}, facilitates
recognition of emission features in the eigenspectra. In this section we add
$\mathbf{P}$ to simulations to investigate if it can be helpful in (1)
distinguishing cross talk and (2) constraining the single emission-line
region model.

In order to mimic real SDSS data, we generate artificial spectra using the
same method described in \S 3.3 of \citet{hu08b}. We add Gaussian random noise
using a realistic noise pattern that is generated by scaling a real error
array taken from SDSS observations to match the desired S/N of the simulation,
and we adopt a realistic mask array. To match our SPCA sample, the redshift
and S/N of the spectra are randomly set to be uniformly distributed as $z \sim
\rm{U}(0.1,0.67)$ and $\rm{S/N} \sim \rm{U}(10,20)$. \feii\ emission is
modeled by Gaussian broadening, scaling, and shifting the template constructed
by \citet{bg92}. All the other emission-line components have a Gaussian
profile. The distributions of EWs are chosen to be consistent with those in
the real SDSS sample.  In each case below, 1000 spectra are generated for
SPCA. Table \ref{tab-simpara} summarizes the models and parameters of each
simulation below. 

\begin{deluxetable*}{cccccccc}
  \tablewidth{0pt}
  \tablecolumns{8}
  \tablecaption{Models and parameters of the simulations
  \label{tab-simpara}}
  \tablehead{
  \colhead{} & \colhead{Power law\tablenotemark{a}} &
  \colhead{\feii\tablenotemark{b}} & \colhead{\oiii\tablenotemark{b}} & 
  \multicolumn{2}{c}{${\rm H}\beta_{\rm BC}^1$} & 
  \colhead{${\rm H}\beta_{\rm BC}^2$\tablenotemark{b}} & \colhead{} 
  \\ \cline{5-6}  
  \colhead{\rb{Case}} & \colhead{$\alpha$} & \colhead{EW} & \colhead{EW} &  
  \colhead{EW} & \colhead{FWHM} & \colhead{EW} & \colhead{\rb{Results}} }
  \startdata
  1(a) & 0 & $\rm{U}(25,75)$ & $\rm{U}(0,20)$ & \nodata & \nodata & \nodata & 
  Fig. \ref{fig-simct}(a)\\
  1(b) & 0 & $\rm{U}(25,75)$ & 20-0.2EW(\feii) & \nodata & \nodata &
  \nodata & Fig. \ref{fig-simct}(b)\\
  1(c) & 0 & $\rm{U}(25,75)$ & ($\rm{U}(0,20)$ + 20-0.2EW(\feii))/2 &
  \nodata & \nodata & \nodata & Fig. \ref{fig-simct}(c)\\
  \hline
  2 & $\rm{U}(-2.5,-0.5)$ & $\rm{U}(25,75)$ & 
  \nodata & $\rm{U}(20,100)$ & $10^{3.35}$EW(\hb)/EW(\feii) & \nodata &
  Fig. \ref{fig-simvm} \\ \hline
  3 & $\rm{U}(-2.5,-0.5)$ & $\rm{U}(25,75)$ & $\rm{U}(0,20)$ &
  0.5EW(\feii) & FWHM(\feii) & $-40\alpha$ & Fig. \ref{fig-simre}
  \enddata
  \tablecomments{\nodata\ means not included in the specific model.}
  \tablenotetext{a}{$f_{5100}$ is fixed to unity.} 
  \tablenotetext{b}{The FWHMs of \feii, \oiii, and ${\rm H}\beta_{\rm BC}^2$
  are 1500, 400, and 4500 \kms, respectively.}
\end{deluxetable*}

\subsection{Cross Talk}
\label{sec-simct}

The aim of the first series of simulations is to study the behavior of the
fractional-contribution spectrum, to see whether it can help distinguish
between cross talk from a genuine inverse correlation when absorption-like
features such as those in eigenspectrum 3 are detected. We begin with the
simplest case, a spectrum consisting of a continuum set to unity and only two
emission-line components, \feii\ and \oiii. The FWHMs of the lines are fixed
to 1500 \kms\ and 400 \kms, respectively, and no shifts are considered. Only
the EWs vary: EW(\feii) $\sim \rm{U}(25,75)$ \AA, and according to how
EW(\oiii) is generated, three different simulations are performed (Table
\ref{tab-simpara}, case 1).
\begin{itemize}
  \item[(a)] The EW of \oiii\ is independent from that of \feii:
\begin{equation}
  {\rm EW}(\oiii) = {\rm EW}_1 \sim {\rm U}(0,20)~\mathring{\rm A}.
\end{equation}
  \item[(b)] The EW of \oiii\ is dependent on that of \feii:
\begin{equation}
  {\rm EW}(\oiii) = {\rm EW}_2 = 20 - 0.2 {\rm EW}({\rm Fe}~\textsc{ii}).
\end{equation}
  \item[(c)] The EW of \oiii\ is semi-dependent on that of \feii:
\begin{equation}
  {\rm EW}(\oiii) = ({\rm EW}_1 + {\rm EW}_2) / 2.
\end{equation}
\end{itemize}

\begin{figure*}
  \centering
  \includegraphics[width=0.98\textwidth]{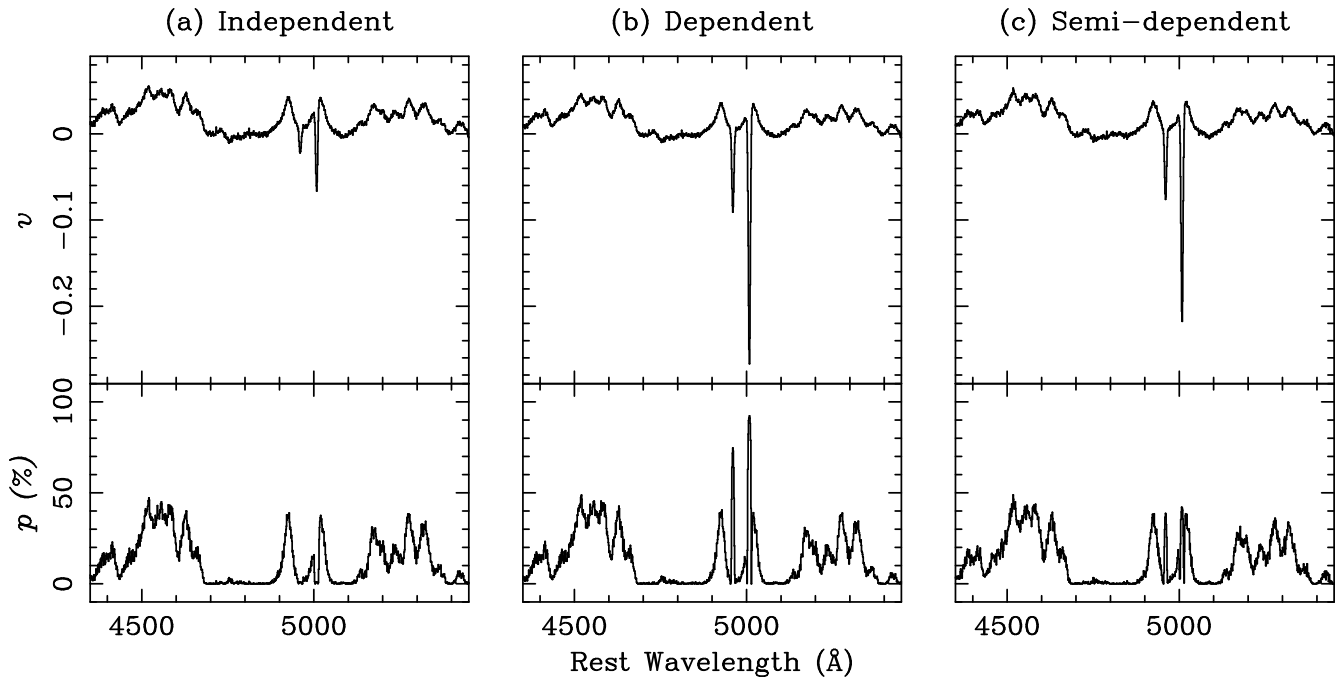}
  \caption{Results of the simulations for cross talk. Panels (a), (b), and (c)
  show the resultant eigenspectra when the EW of \oiii\ depends differently on
  the EW of \feii, as indicated by the labels above the panels. Note the
  different fractional contribution ($p$) at the wavelength of \oiii.}
  \label{fig-simct}
\end{figure*}

Figure \ref{fig-simct} shows the results of the three simulations. The three
eigenspectra all show positive \feii\ emission and negative \oiii\ lines.
Although the strengths of the negative \oiii\ features relative to \feii\ are
different in the three cases, nothing essential can be distinguished. But the
three fractional-contribution spectra show remarkable differences at the
positions of \oiii\ lines: zero in case (a), close to 100\% in case (b), and
$\sim$ 40\% in case (c). Thus the fractional-contribution spectrum, which
describes the contribution of eigenspectra to emission features, effectively
distinguishes between cross talk and real correlation. 

\begin{figure}
  \centering
  \includegraphics[width=0.45\textwidth]{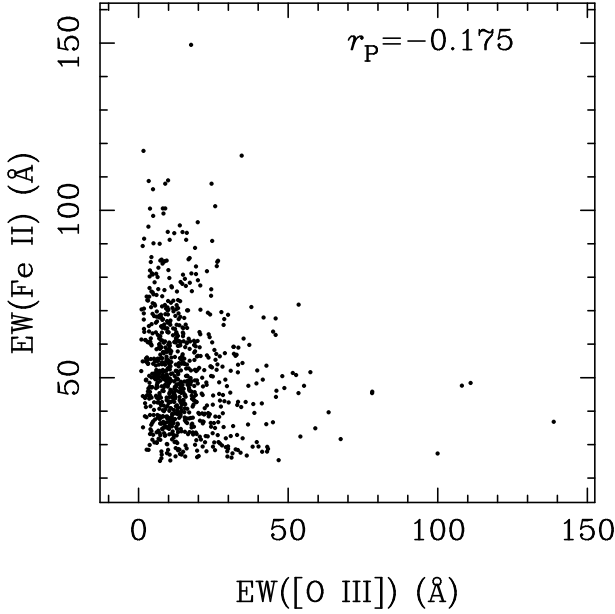}
  \caption{EW of \feii\ versus EW of \oiii\ for our sample used for SPCA
  analysis. The lack of a strong correlation suggests that the negative narrow
  \oiii\ feature in our eigenspectrum 3 is not caused by BG92's EV1.}
  \label{fig-ewo3fe}
\end{figure}

Eigenspectrum 3 of our sample (Figure \ref{fig-esall}(d)) resembles case (a)
here. The fractional-contribution spectrum goes almost to zero for \oiii,
meaning that the negative \oiii\ features arise only from cross talk and do
not reflect any intrinsic inverse correlation between \feii\ and \oiii.  But
where is BG92's EV1 in our eigenspectra? We believe that this signature is
absent from our sample because of selection effects. By design, our sample is
biased toward objects with strong \feii\ (\S \ref{sec-sample}), and hence we
do not expect to find a strong inverse correlation between the strengths of
\feii\ and \oiii. Figure \ref{fig-ewo3fe} confirms this expectation; EW(\feii)
and EW(\oiii) are only correlated at the level of $r_{\rm P}=-0.175$. This
correlation is much weaker than those between Balmer lines, narrow emission
lines, or \feii\ lines, thus will not appear in the first three eigenspectra.

\subsection{The Single \hb\ Component Model}
\label{sec-simvw}

\begin{figure}
  \centering
  \includegraphics[width=0.45\textwidth]{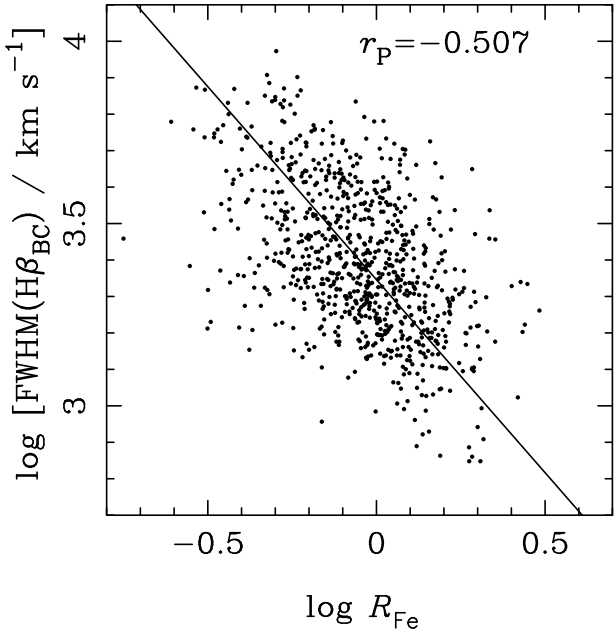}
  \caption{Correlation diagram for FWHM(\hbb) versus \rfe. The solid line is
  the fit to the data as given in Equation (\ref{eqn-hbwrfe}).}
  \label{fig-hbwrfe}
\end{figure}

Before describing the second simulation, which is aimed at ruling out the
single Balmer line component model, we plot the correlation diagram for
FWHM(\hbb) versus \rfe, defined as the ratio between the EW of \feii\ and the
EW of \hbb\ (Figure \ref{fig-hbwrfe}).  There is a rather strong inverse
correlation ($r_{\rm P} = -0.507$), which, from an OLS bisector fit, can be
described by
\begin{equation}
  \label{eqn-hbwrfe}
  {\rm FWHM}({\rm H}\beta_{\rm BC}) = 10^{(3.35\pm0.01)} ~
  {R_{\rm Fe}}^{(-1.05\pm0.02)}.
\end{equation}
This correlation has been found before \citep[e.g.,][and references
therein]{bg92,sulentic00a}, but its physical interpretation is still far from
conclusive. There are two possible phenomenological interpretations. The
straightforward explanation is that there is only a single \hb\ component
whose width varies systematically with the relative strength of \feii\ and
\hb. Alternatively, \hb\ has two kinematic components, and the intensity of
the narrower one scales with that of \feii\ by some fixed ratio, while the
broader one varies independently. In this scenario, larger \rfe\ means that
\hb\ has a stronger narrow component relative to its broader component, making
the whole \hb\ profile narrower. Previous SPCA studies have claimed that
eigenspectra such as our eigenspectra 1 and 3 can be produced by either of the
models above. There was no definitive way to discriminate between the two.

Here we aim to use our SPCA results to distinguish between these two
phenomenological models, with the help of the fractional-contribution spectra
$\mathbf{P}$. We generate artificial spectra that consist of three emission
components (Table \ref{tab-simpara}, case 2):
\begin{itemize}
  \item[(1)] \feii\ emission whose EW $\sim \rm{U}(25,75)$ \AA;
  \item[(2)] an \hb\ component whose EW $\sim \rm{U}(20,100)$ \AA\ and FWHM $=
    10^{3.35}{R_{\rm Fe}}^{-1}$, according to Equation (\ref{eqn-hbwrfe});
  \item[(3)] a power-law continuum whose $f_{5100}$ flux density is fixed to
    unity but slope varies in the range $\alpha \sim \rm{U}(-2.5,-0.5)$.
\end{itemize}
A varying power law is added here to mimic real spectra more realistically
than in the first series of simulations in the last subsection, and it helps
us achieve our goal, as shown below.

\begin{figure*}
  \centering
  \includegraphics[width=0.9\textwidth]{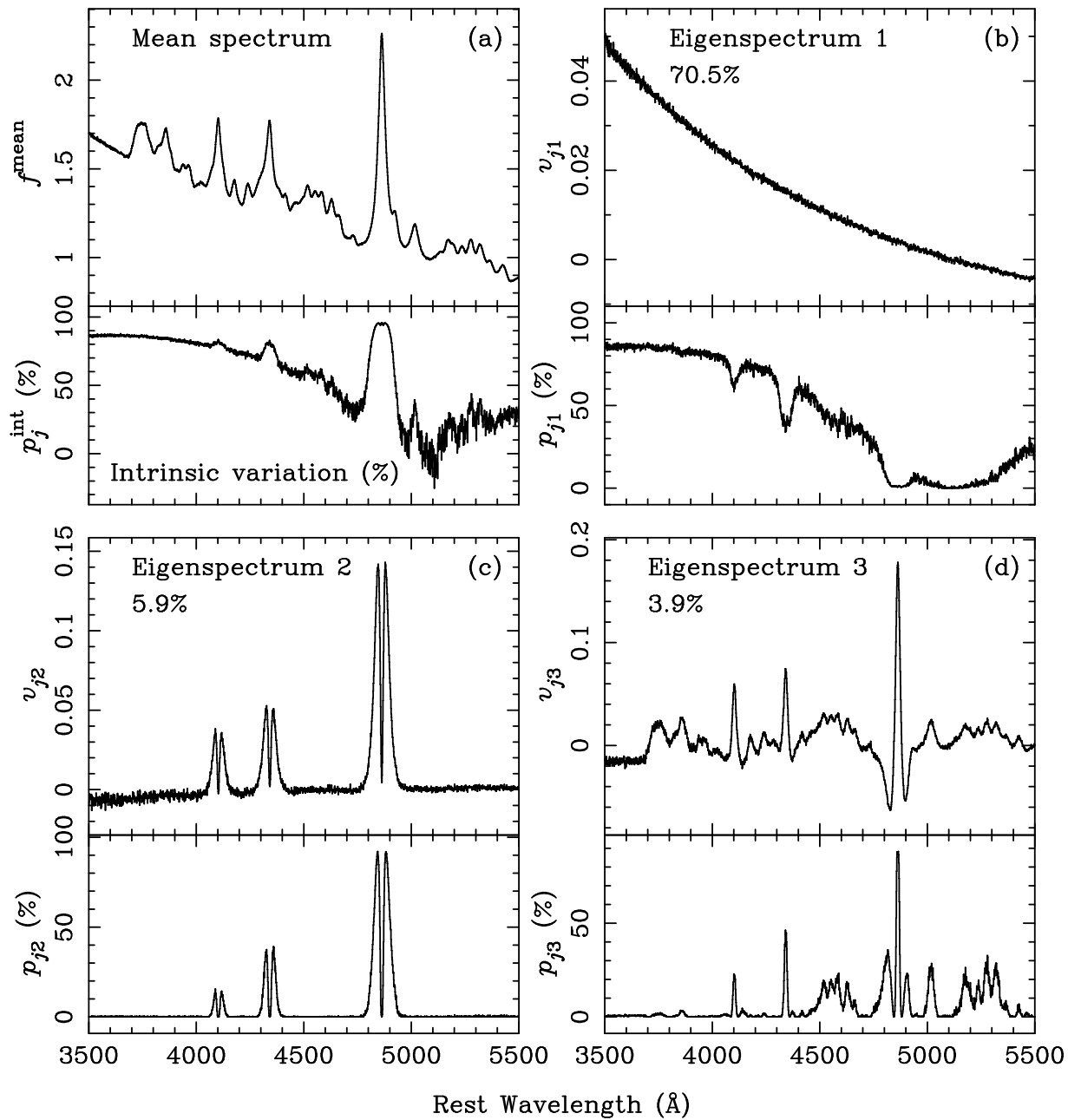}
  \caption{Results of the SPCA simulation for a single-component \hb\ line
  whose width correlates with \rfe. The labels are the same as those in Figure
  \ref{fig-esall}. While the mean spectrum resembles that of Figure
  \ref{fig-esall}, the first three eigenspectra and their corresponding
  fractional-contribution spectra differ. Note that the simulation here does
  not take into account the \oiii\ lines. See text for details.}
  \label{fig-simvm}
\end{figure*}

Figure \ref{fig-simvm} shows the mean spectra and the first three eigenspectra
given by SPCA. As in Figure \ref{fig-esall}, the lower panels show the
proportion of intrinsic variation and the fractional-contribution spectra.
Considering that we do not account for the \oiii\ lines in the simulation, the
mean spectra and proportion of intrinsic variation here resembles that in
Figure \ref{fig-esall}. There are three main differences between these SPCA
results based on simulated spectra compared to those derived from real
spectra. First, the eigenspectrum 1 here shows only a clean power law; some
emission lines that correlate with the continuum are missing. Second,
eigenspectrum 2 highlights the wings of the Balmer lines. And third,
eigenspectrum 3, which contains \feii\ emission, shows a ``W'' shape at \hb,
and, as shown by the fractional-contribution spectrum, contributes a lot to
the variation of \hb\ wings. 

In conclusion, a single Balmer emission-line component whose width simply
correlates with its strength with respective to \feii\ generates simulated
eigenspectra that do not match well with our observed results in any of the
first three eigenspectra and fractional-contribution spectra.

\subsection{Reproduction of Our SPCA Results}
\label{sec-simre}

The simulation above suggests that neither a single Balmer emission-line
component nor two independently varying components can produce our SPCA
results shown in Figure \ref{fig-esall}. A Balmer line component that
correlates with the power-law continuum plus another that correlates with
\feii\ emission are needed.  We demonstrate this using the simulation below.

As listed on the last line of Table \ref{tab-simpara}, the artificial spectra
are composed of the following five components:
\begin{itemize}
  \item[(1)] a power-law continuum with specific flux density $f_{5100}$ 
    fixed to unity but slope $\alpha \sim \rm{U}(-2.5,-0.5)$;
  \item[(2)] \feii\ emission with EW $\sim \rm{U}(25,75)$ \AA\ and a fixed
    FWHM = 1500 \kms;
  \item[(3)] \oiii\ emission with EW $\sim \rm{U}(0,20)$ \AA\  and a fixed 
   FWHM = 400 \kms;
  \item[(4)] a Balmer component with EW set to $0.5$EW(\feii) and FWHM fixed 
    to that of \feii;
  \item[(5)] a broader Balmer component with EW $= -40\alpha$ and a fixed 
    FWHM $= 4500$ \kms.
\end{itemize}
The fourth component depends on the second, and the fifth is determined by the
first. Thus, there are only three free parameters: the slope of the power-law
continuum, the strength of the \feii\ emission, and the strength of the narrow
\oiii\ lines.

\begin{figure*}
  \centering
  \includegraphics[width=0.9\textwidth]{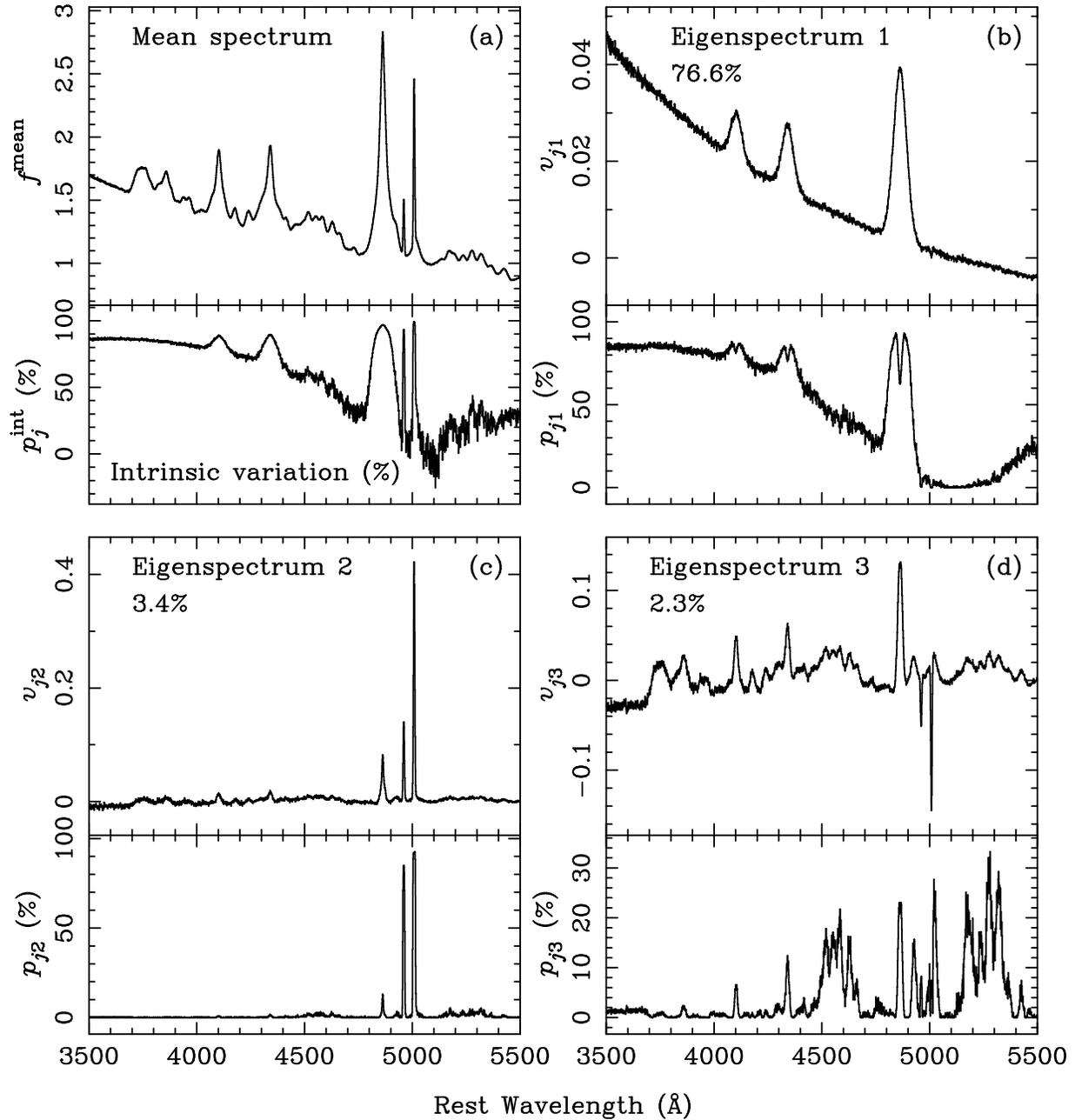}
  \caption{Results of the SPCA simulation aimed at reproducing the
  eigenspectra of the observed sample.  The labels are the same as those in
  Figure \ref{fig-esall}. The first three eigenspectra are sufficient to
  account for the intrinsic variation. Ignoring the emission features that are
  not included in the simulation (e.g., \oii\ emission, higher order Balmer
  lines, Balmer continuum, etc.), the mean spectrum, the eigenspectra, and the
  fractional-contribution spectra closely resemble their observed counterparts
  in Figure \ref{fig-esall}.}
  \label{fig-simre}
\end{figure*}

Figure \ref{fig-simre} shows the derived SPCA results. The first three
eigenspectra, added up, account for 82.40\% of the total variation, and are
sufficient for describing the intrinsic variation (82.26\% of the total). This
is consistent with the number of free parameters included in the simulation.
Comparing the results here with those in Figure \ref{fig-esall}, this
simulation successfully reproduces the SPCA results for the real SDSS spectra,
including almost all of the main features in the mean spectrum, the first
three eigenspectra, and the fractional-contribution spectra.

\vspace{1em}

These simulations strongly confirm the validity of the fractional-contribution
spectrum and verify our interpretation of the eigenspectra in \S
\ref{sec-pcaresult}. They support the notion that two Balmer emission-line
components are needed to reproduce our SPCA results: (1) a broader one that
correlates with the power-law continuum and produces no \feii\ emission and
(2) a narrower one that has the profile of, and correlates with, \feii.

\section{Results for Large \feii\ Velocity Shifts}
\label{sec-ovbin}

Aside from the primary sample of quasars with small \feii\ velocity shift,
four other samples with larger \feii\ velocity shifts are also established in
\S \ref{sec-sample}. This section presents the results of SPCA for these four
samples, which can be understood easily in light of the analysis and
simulations for the primary sample above.

\begin{figure*}
  \centering
  \includegraphics[width=0.9\textwidth]{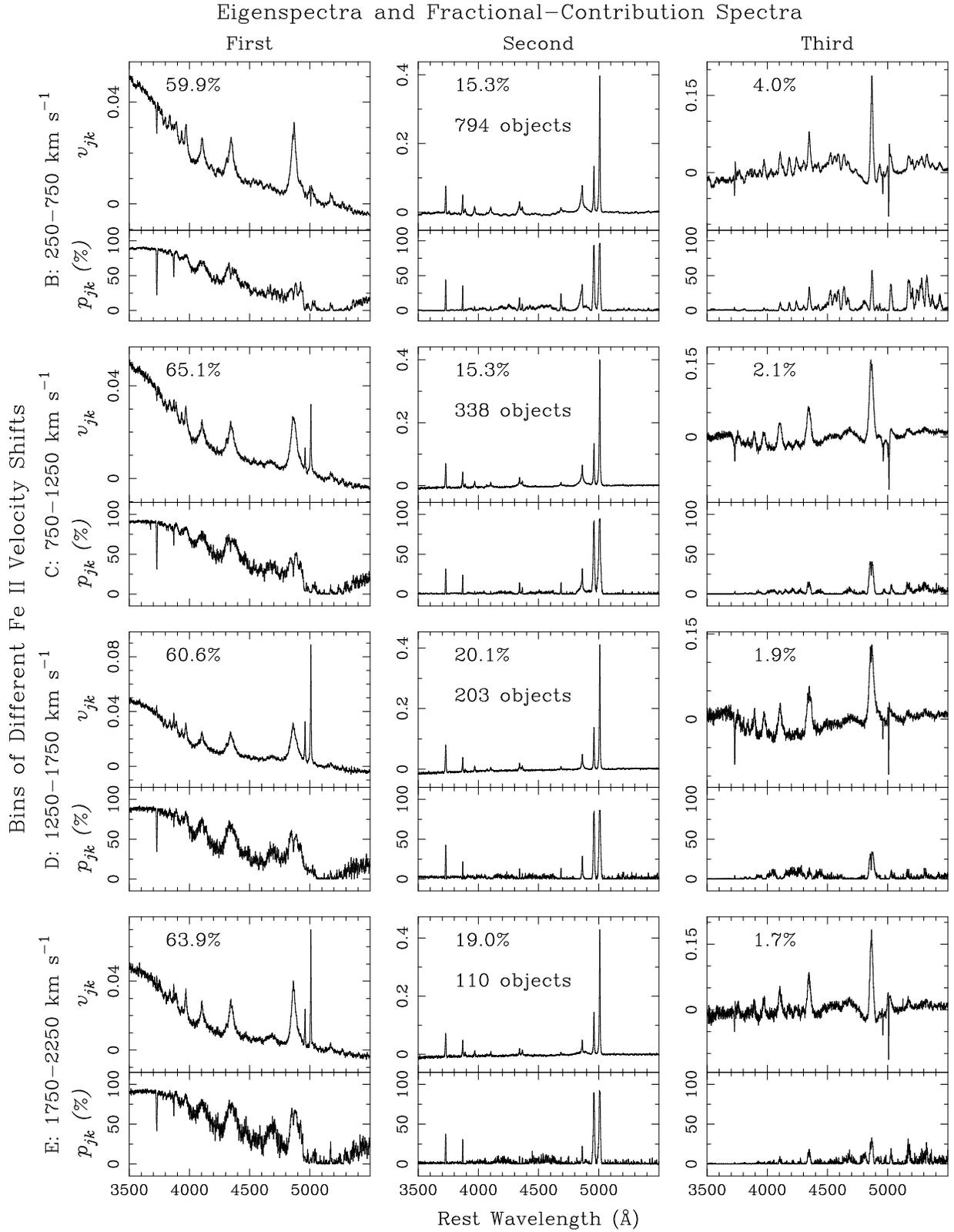}
  \caption{The first three eigenspectra and corresponding
  fractional-contribution spectra for the four bins of large \feii\ velocity
  shift. The range of \feii\ velocity shift in each bin is labeled on the left
  of each row. The percentage in each panel indicates the contribution of the
  eigenspectrum over the entire wavelength range. The number of objects in
  each of the four bins is given in the panels of the middle column.}
  \label{fig-esvbin}
\end{figure*}

Figure \ref{fig-esvbin} shows the first three eigenspectra and corresponding
fractional-contribution spectra for the four new velocity bins. Sample B
(with \feii\ velocity shifts between 250 and 750 \kms) has roughly equal size
(794 objects) as the primary sample. The results of SPCA for this sample
resemble those for the primary sample, for example, in terms of the
contributions of the eigenspectra over the entire wavelength range (59.9\%,
15.3\%, and 4.0\%), the shapes and intensities of the first three
eigenspectra, and fractional-contribution spectra. Thus, the first three
eigenspectra represent the same emission components described in \S
\ref{sec-pcades}.

\begin{figure}
  \centering
  \includegraphics[width=0.45\textwidth]{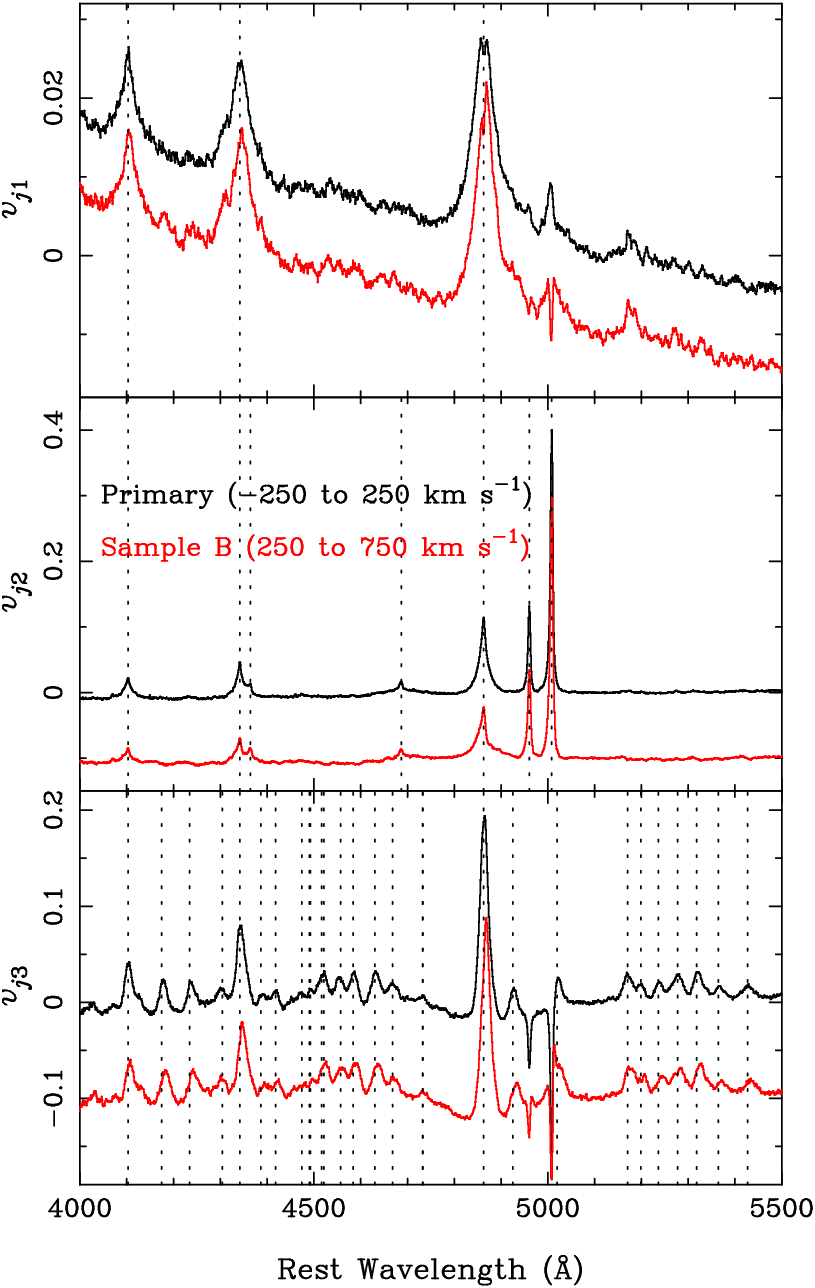}
  \caption{Comparison between the first three eigenspectra of the primary
  sample (in black) and sample B (in red). The eigenspectra of sample B are
  shifted vertically for clarity. The dotted lines mark the rest wavelengths
  of major emission lines encoded in the eigenspectra (top panel: Balmer
  lines; middle panel: Balmer lines, \heii\ $\lambda$4686, \oiii\ $\lambda$
  4363, and \oiii\ $\lambda\lambda$4959, 5007; bottom panel: Balmer lines, and
  \feii\ lines). Note the redshift of eigenspectrum 3 of sample B.}
  \label{fig-escomp}
\end{figure}

In Figure \ref{fig-escomp}, we compare the first three eigenspectra of sample
B (in red, vertically shifted) to those of the primary sample (in black). The
rest wavelengths of major emission lines are marked by dotted lines. Top and
middle panels show that eigenspectra 1 and 2 of the two samples are almost the
same. None of the major emission lines (very broad Balmer lines in
eigenspectrum 1; narrow Balmer lines, \heii, and \oiii\ lines in eigenspectrum
2) have a significant velocity shift between the two samples. The only notable
difference is seen in eigenspectrum 3. The \feii\ lines and the
intermediate-width Balmer lines of sample B are redshifted by $\sim 400$ \kms\
with respect to the primary sample. This is expected because the two samples
are divided by \feii\ velocity shift and the third eigenspectrum represents
the intermediate-width emission lines, which mainly include \feii.
It provides additional evidence that the measurement of \feii\ velocity shift
in \citet{hu08b} is reliable in these two samples.
Note that the negative features of \oiii\ arise from cross talk, not redshift.

The three bins with larger \feii\ velocity shift contain relatively fewer
objects: 338, 203, and 110 in samples C, D, and E, respectively. The
eigenspectra and fractional-contribution spectra of these three samples are
very similar. Their first two eigenspectra and fractional-contribution spectra
resemble those of the primary sample (judging from the fractional-contribution
spectra, note that the \oiii\ lines in eigenspectra 1 arise just from cross
talk), while fractional-contribution spectra 3 are different. Eigenspectra 3
of these three samples contribute little in terms of \feii\ emission. Their
contributions over the entire wavelength range are also minimal. Thus,
eigenspectra 3 of samples C, D, and E do not record the intermediate-width
component. This is reasonable because EW(\feii) in the three samples is small
and \hb\ is rather broad \citep{hu08b} and tends to have single-Gaussian
profiles (Figure 1(a) of \citealt{hu08a}). The intermediate-width component is
expected to be weak, contributes little to the total variation, and can be
easily smeared by nonlinear factors. This interpretation is supported by the
fact that the first two eigenspectra have already contributed a larger
proportion of variation ($\gtrsim$ 80\%) than the first three eigenspectra of
the primary sample (78.4\%).

In conclusion, the results of SPCA for sample B resemble those for the primary
sample, except that eigenspectrum 3 is redshifted. For the other three
samples, only the first two eigenspectra are important; they record the
power-law continuum $+$ very broad emission lines and narrow emission lines,
respectively.

\section{Discussion}
\label{sec-dis}

Our SPCA analysis is based on the quasar sample of \citet{hu08b}, which is
intentionally biased toward objects with strong \feii\ emission [EW(\feii) $>
25$ \AA] to facilitate measurement of \feii\ velocity shifts. Given this
situation, it is important to ask whether the principal result of this paper,
namely that the Balmer lines contain two physically distinct components, is
unique to our sample or applies to the quasar population in general. The key
test of this scenario is to perform a detailed decomposition of the broad \hb\
profile, to isolate the two kinematic components. We defer this analysis to
the second paper of this series, where we will explore the feasibility of
using the eigenspectrum 3 derived here as a template to model the
intermediate-velocity component of the spectrum.

In the meantime, we will use two samples selected differently from that used
in our SPCA analysis to argue that our main conclusions are robust against
sample selection effects.

\subsection{Intermediate-line Region}

Our analysis posits that the traditional BLR consists of two components, one
with intermediate velocities in which \hb\ and \feii\ coexist with roughly
constant relative intensity, and another characterized by larger velocities
that emits \hb\ but no \feii. The relative contribution of these two
components varies from source to source. A simple consequence of this picture
is that the EW of the total broad \hb\ line should be larger than a factor
times that of \feii.  We construct a sample from the SDSS DR5 quasar catalog
using the same criteria of \citet{hu08b}, as described in \S\ref{sec-sample},
except that we impose no restriction based on EW(\feii). Thus, this sample is
not biased toward objects with strong \feii\ emission.

\begin{figure}
  \centering
  \includegraphics[width=0.45\textwidth]{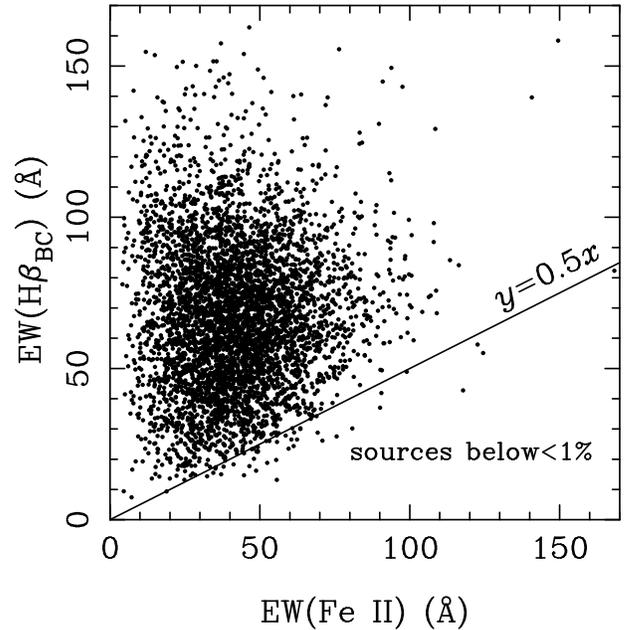}
  \caption{EW of \hbb\ versus that of \feii. This sample is selected from the
  SDSS DR5 quasar catalog by the same criteria of those in \citet{hu08b},
  except that no cut is made for EW(\feii). The solid line denotes EW(\hbb) $=
  0.5$ EW(\feii). Fewer than 1\% of the sources lie below this line.}
  \label{fig-hbfeew}
\end{figure}

Figure \ref{fig-hbfeew} shows the relation between \hbb\ and \feii\ EWs for
this sample. As in Figure \ref{fig-corall}, we measure \hbb\ by Gauss-Hermite
fitting, which accounts for the full broad \hb\ profile. The distribution of
points in the plot shows a clear lower boundary, which is approximately
delineated by EW(\hbb) $= 0.5$ EW(\feii). Only 47 sources out of 4757 (fewer
than 1\%) lie below the line. The existence of this bottom boundary is
consistent with our two-component scenario if the strength of the
intermediate-width \hb\ component is roughly a fixed fraction of \feii.
Furthermore, our simulations confirm that such an assumption reproduces well
the eigenspectra of the real data.

\subsection{Correlation Between Strength of Very Broad \hb\ and $\alpha$}
\label{sec-ewalpha}

A correlation between the EW of broad \hb\ and the optical continuum slope
$\alpha$ has seldom been reported in the literature, and the few papers that
have mentioned it have given inconsistent results. \citet{srianand97}, for
example, found no correlation between optical spectral index and the EW of
\hb, but \citet{francis01} reported that the EW of \hb\ increases as the
continuum becomes bluer. Our finding that the EW of very broad \hb\ correlates
with $\alpha$ does not necessarily contradict with previous studies because no
\hb\ decomposition was done before, not to mention that the samples are
different. \citet{richards03} found that \hb\ line width decreases in quasars
with redder continuum; this trend is qualitatively consistent with our
scenario insofar as bluer objects exhibit a stronger very broad \hb\ component
and hence a broader overall line profile.

The sources with EW(\feii) $< 25$ \AA\ in Figure \ref{fig-hbfeew} comprise
quasars with weak \feii\ emission; they formally lie outside of the selection
criterion of \citet{hu08b}. Within the framework of this study, this subsample
of weak-\feii\ sources should have a weak intermediate-width \hb\ component,
and their broad \hb\ emission should be dominated by the very broad component.
This subsample is convenient for testing the inverse correlation between the
strength of the very broad \hb\ component and the slope of the power-law
continuum, without having to perform accurate profile decomposition.

\begin{figure}
  \centering
  \includegraphics[width=0.45\textwidth]{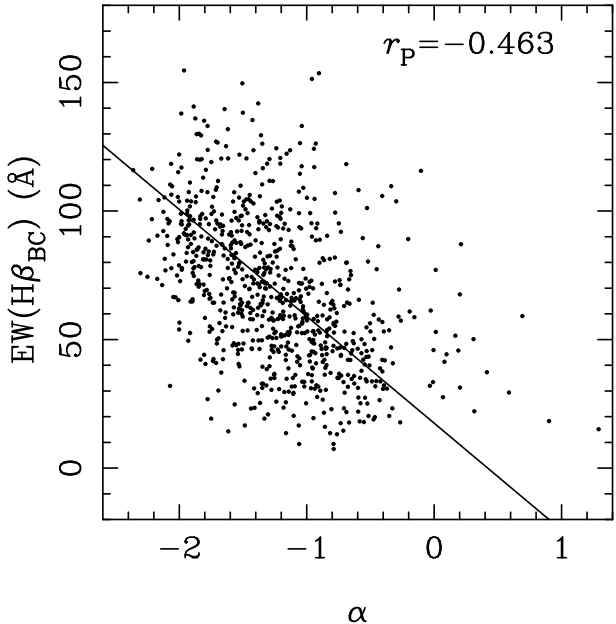}
  \caption{Correlation diagram for the EW of entire broad \hb\ line versus the
  slope of the power-law continuum ($\alpha$, defined as $f_\lambda \propto
  \lambda^\alpha$) for sources with EW(\feii) $< 25$ \AA\ in Figure
  \ref{fig-hbfeew}. The inverse correlation is strong. The solid line shows
  the OLS bisector fit.}
  \label{fig-alphahbew}
\end{figure}

Figure \ref{fig-alphahbew} confirms our expectations. We find a strong inverse
correlation between EW(\hbb) and $\alpha$, with Pearson's correlation
coefficient $r_{\rm P}=-0.463$. An OLS bisector fit yields 
\begin{equation}
  {\rm EW}({\rm H}\beta_{\rm BC}) = (17.5\pm2.7) - (41.6\pm1.3)\alpha.
\end{equation}
This empirical result motivated one of the input criteria (the fifth) for the
simulation described in \S \ref{sec-simre}.

A bluer continuum has more ionizing photons relative to the non-ionizing
continuum at 5100 \AA, to which the EWs in the present paper all refer. Thus,
an inverse correlation between the EW of an emission line and $\alpha$ arises
naturally if the line-emitting clouds are optically thick to, and photoionized
by, the same continuum that we see, and the clouds emits isotropically.  On
the other hand, the lack of any correlation between the intermediate-width
component and the continuum slope reflects the complexity of the dynamics
\citep{hu08b} and physics \citep[e.g.,][and reference
therein]{baldwin04,ferland09} of the \feii-producing gas. These two different
dependences on spectral slope, in conjunction with the distinct locations and
dynamics implied by their different velocity widths and shifts, indicate that
the two components may respond very differently to continuum variations. This
has been suggested by the reverberation analysis of \feii\ emission in Ark 120
\citep{kuehn08}, and also by recent studies that investigate the time delay
between different velocity components of \hb\ and the continuum
\citep{zhang11,wangj11}. Velocity-resolved time-delay measurements in some
sources also show that \hb\ originates from structures more complicated than a
single virialized region \citep[e.g.,][and reference
therein]{bentz10,denney10}.

The width of the broad \hb\ emission line has been widely used to estimate
virial velocities to derive BH masses in AGNs. Our two-component model for the
\hb-emitting region raises an important question: which component better
traces the virialized portion of the BLR? Our SPCA results suggest that the
very broad component is more appropriate for BH mass estimation. We will
investigate this problem in the third paper of this series.

\section{Summary}
\label{sec-sum}

We select a sample of 816 quasars with small \feii\ velocity shift from SDSS
and perform spectral principal component analysis (SPCA) on this sample in
rest wavelength range 3500--5500 \AA. Apart from adjusting some details in the
SPCA algorithm, we introduce, for the first time, a parameter called \emph{the
fractional-contribution spectrum} that measures \emph{the proportion of the
variation accounted for by an eigenspectrum in a wavelength bin}. We
demonstrate the utility of this new parameter in helping to interpret the
eigenspectra. We explore the correlations between the weights of the
eigenspectra and various physical quantities to confirm the physical meaning
of the eigenspectra. We perform a bootstrap analysis, spectral fitting of the
eigenspectra, and Monte Carlo simulations to test the uncertainty of the
eigenspectra, the validity of the fractional-contribution spectrum we
introduced, and the two-component model of the \hb-emitting region we propose.

Our principal findings from SPCA are as follows.

\vspace{1em}

1. The first three eigenspectra have clear physical meanings. Eigenspectrum 1
represents the power-law continuum, very broad Balmer emission lines, and the
correlation between them. Eigenspectrum 2 represents narrow emission lines.
Eigenspectra 3 consists of \feii\ lines and Balmer lines with kinematically
similar intermediate velocities.

2. The fractional-contribution spectrum is a powerful tool for diagnosing the
emission features represented in each eigenspectrum, as well as for
recognizing spurious features that arise from cross talk.

3. The broad \hb\ line consists of two physically distinct components: a very
broad component whose strength correlates with the slope of the optical
power-law continuum, and an intermediate-width component that has the profile
of, and correlates with, \feii.

\vspace{1em}

While our current sample is biased toward objects with strong \feii\ emission,
we argue that our overall findings are immune from strong sample selection
bias and are generally applicable to the general quasar population. This
means that the strength of the intermediate-width \hb\ component varies from
source to source and could vanish. In extreme cases \hb\ only has a single
Gaussian component. We will rigorously test this conclusion in a forthcoming
work, where we will perform detailed spectral decomposition using the new
template spectrum for \feii\ emission and intermediate-velocity Balmer lines
derived from eigenspectrum 3.  

The two-component nature of the broad \hb\ line raises important concerns
about the robustness of current virial BH mass estimates that make use of the
\hb\ line width. This issue will be explored in another forthcoming
publication.

\acknowledgments

We are grateful to the referee for his/her careful, detailed comments and
suggestions that helped to improve the paper. CH is grateful to Zhao-Hui
Shang and Wei-Hao Bian for discussions. We appreciate extensive discussions
among the members of IHEP AGN group. This research is supported by NSFC via
NSFC-10903008, 11133006, 11173023, 11233003, and 973 Program 2009CB824800. The
work of LCH is funded by the Carnegie Institution for Science. GJF
acknowledges support by NSF (1108928, and 1109061), NASA (10-ATP10-0053,
10-ADAP10-0073, and NNX12AH73G), and STScI (HST-AR-12125.01, GO-12560, and
HST-GO-12309).

This paper has used data from the SDSS.
Funding for the SDSS and SDSS-II has been provided by the Alfred P. Sloan
Foundation, the Participating Institutions, the National Science Foundation,
the U.S. Department of Energy, the National Aeronautics and Space
Administration, the Japanese Monbukagakusho, the Max Planck Society, and the
Higher Education Funding Council for England. The SDSS Web Site is \url{{\tt
http://www.sdss.org/}}.

The SDSS is managed by the Astrophysical Research Consortium for the
Participating Institutions. The Participating Institutions are the American
Museum of Natural History, Astrophysical Institute Potsdam, University of
Basel, University of Cambridge, Case Western Reserve University, University of
Chicago, Drexel University, Fermilab, the Institute for Advanced Study, the
Japan Participation Group, Johns Hopkins University, the Joint Institute for
Nuclear Astrophysics, the Kavli Institute for Particle Astrophysics and
Cosmology, the Korean Scientist Group, the Chinese Academy of Sciences
(LAMOST), Los Alamos National Laboratory, the Max-Planck-Institute for
Astronomy (MPIA), the Max-Planck-Institute for Astrophysics (MPA), New Mexico
State University, Ohio State University, University of Pittsburgh, University
of Portsmouth, Princeton University, the United States Naval Observatory, and
the University of Washington. 

\appendix

\section{Evidence for Redshifted \feii\ Emission in Quasars from Composite
Spectra}
\label{app-feschi}

\citet{sulentic12} called into question the measurement of \feii\ velocity
shifts (\vfe) in \citet{hu08b}. They raised two criticisms. First, they argue
that the majority of the sources in \citet{hu08b} do not have enough S/N to
yield a reliable measurement of \vfe. Second, they note that \citet{hu08b} did
not include \heii\ emission in their fits. \citet{sulentic12} generated
composite spectra, which have high S/N, of several subsamples defined by their
4D Eigenvector 1 formalism, and then measured \vfe\ from fits of the composite
spectra that include the \heii\ line. They concluded that the \feii\ emission
in their composite spectra do not have significant velocity shifts. In this
Appendix, we adopt similar fitting methods, take \heii\ into consideration,
measure \vfe\ for the five composite spectra generated in \citet{hu08b}, and
test the statistical significance of our measurement of \vfe. We confirm that
our redshift measurements of \feii\ are robust.

\begin{figure*}
  \centering
  \includegraphics[width=0.85\textwidth]{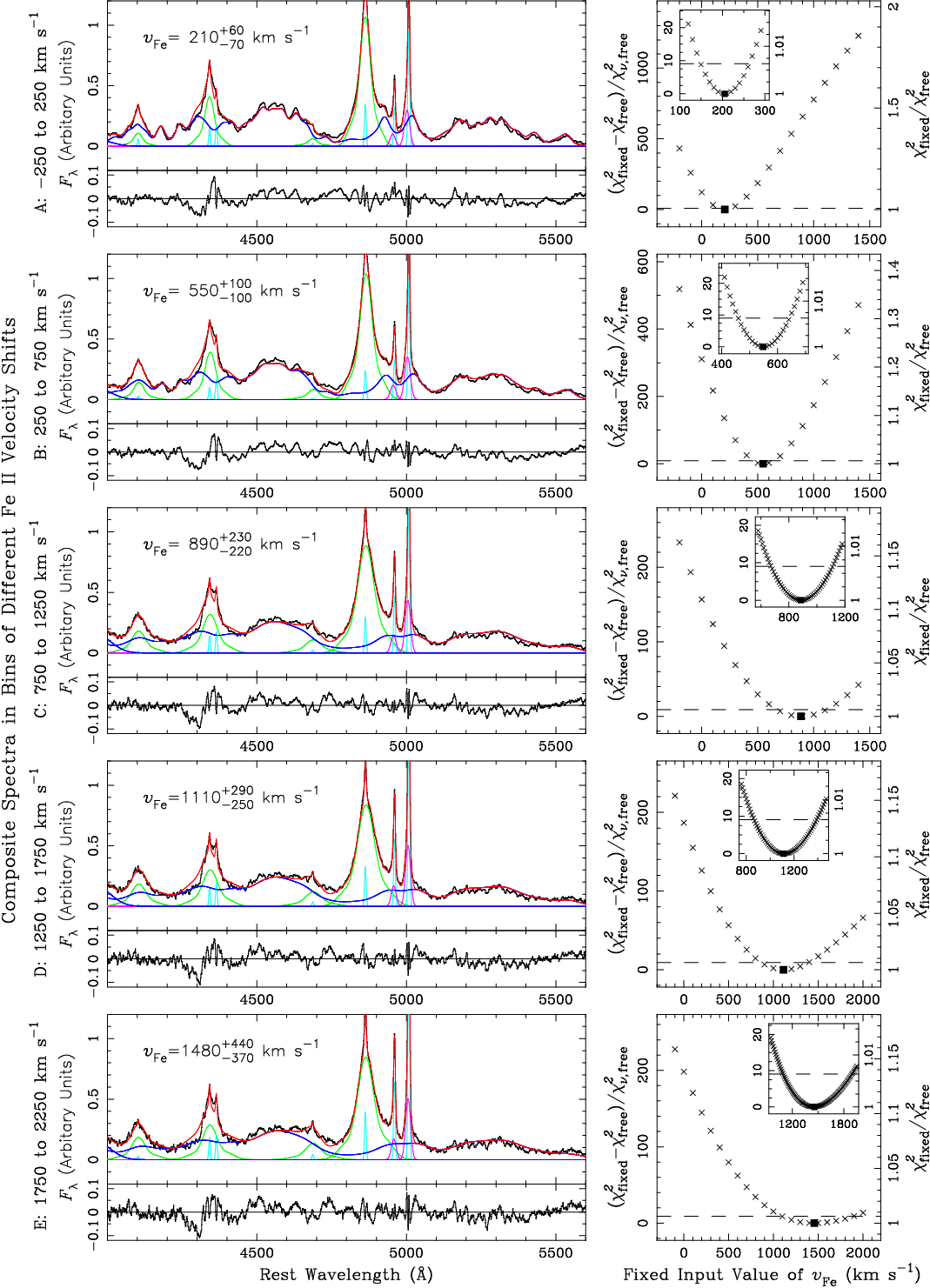}
  \caption{(Left) Fits of composite spectra made by assuming \vfe\ is free.
  The composite spectra were generated in \citet{hu08b} by stacking the
  spectra of quasars in bins of different measured \vfe. The range in each bin
  is labeled on the left of each row. For each bin, the top panel shows the
  continuum-subtracted composite spectrum (black) and best-fit model (red),
  including \feii\ emission (blue), broad emission lines (green), narrow
  emission lines (cyan), and wings of \oiii\ lines (magenta). The bottom panel
  shows the residuals. (Right) \ks\ curve (crosses) for the fits with fixed
  \vfe; the insert zooms in around the minimum. The horizontal dashed line
  marks where the difference in \ks\ has 3$\sigma$ significance. The value of
  the best-fit \vfe\ (solid square) and its 3$\sigma$ confidence interval
  (where the \ks\ curve intersects the dashed line) are labeled in the panel
  showing the fit. Note that the range of the abscissa of the \ks\ curves for
  the last two composite spectra are different with those for the first three.
  }
  \label{fig-feschi}
\end{figure*}

The details of the five composite spectra are described in \S 4.6 of
\citet{hu08b}. Briefly, the composite spectra are geometric means
\citep{vandenberk01} of the spectra of quasars in five bins of different \vfe\
measured in \citet{hu08b}: $-$250 to 250 \kms\ (A, 1350 objects), 250 to 750
\kms\ (B, 1362 objects), 750 to 1250 \kms\ (C, 590 objects), 1250 to 1750
\kms\ (D, 332 objects), and 1750 to 2250 \kms\ (E, 180 objects). Our fitting
method here resembles that in \citet{hu08b} but is improved in two aspects:
(1) the \heii\ line is included and (2) the power-law continuum, \feii\
emission, and other lines are fitted simultaneously. The left column of Figure
\ref{fig-feschi} shows our results. For each velocity shift bin, the top panel
shows the continuum-subtracted composite spectrum (black) and best-fit model
(red). Besides the power-law continuum, the model contains the following
components: (1) \feii\ emission (blue) modeled by broadening, scaling, and
shifting the I Zw 1 Fe template constructed by \citet{bg92}, (2) broad
emission lines (green) modeled by a set of Gaussian-Hermite functions
representing \hb\ $\lambda$4861, \hg\ $\lambda$4340, \hd\ $\lambda$4102, and
\heii\ $\lambda$4686, (3) narrow emission lines (cyan) modeled by a set of
single Gaussians representing \hb, \hg, \hd, \oiii\ $\lambda\lambda$4959,
5007, \oiii\ $\lambda$4363, and \heii\ $\lambda$4686, and (4) wings of the
\oiii\ lines (magenta) modeled by a set of single Gaussians. The lines in each
set have the same shape and shift, but different intensities. The bottom panel
shows the residuals. The fitting is performed in the wavelength window
4000--5600 \AA.

The significance of the measured \vfe\ can be tested by using a \ks\ statistic
to determine whether the model with varying \vfe\ fits the data better than
the model with \vfe\ fixed to a specific value (with other parameters left
free). We use the $F$-test described in \S 11.4 of \citet{bevington03}. In the
fitting described above, if \vfe\ is free to vary the best fit has chi-square
$\chi^2_{\rm free}$ with 1435 degrees of freedom, and the reduced chi-square
$\chi^2_{\nu,{\rm free}} = \chi^2_{\rm free} / 1435$. If we fix \vfe\ to a
constant but allow all other parameters to vary and fit the composite spectra
again, the new best fit has chi-square $\chi^2_{\rm fixed}$ with 1435$+$1
degrees of freedom. From Equation 11.50 of \citet{bevington03}, the quantity
\begin{equation}
  F_\chi = (\chi^2_{\rm fixed}-\chi^2_{\rm free})/\chi^2_{\nu,{\rm free}}
\end{equation}
follows a $F_{1, \nu_2}$ distribution, where $\nu_2 = 1435$ in this case.
Thus, the probability that the model with varying \vfe\ does not improve the
fit compared to the model with \vfe\ fixed to a specific value equals the
probability of exceeding $F_\chi$ in an $F_{1, \nu_2}$ distribution, namely
$P_F(F_\chi; 1, 1435)$. From Figure C.5 of \citet{bevington03}, the value of
$F_\chi$ for $P_F(F_\chi; 1, 1435) = 1 - 99.73\%$ is approximately 9 (the
$\nu_2 = \infty$ curve); the precise number is 9.03.%
\footnote{
This precise value of $F_\chi$ for $P_F(F_\chi; 1, 1435) = 1 - 99.73\%$ can be
obtained using the software {\it R} \citep{r12} by the command {\tt qf(0.9973,
1, 1435)}.
}
Thus, the best-fit value of \vfe\ is significant at more than 3$\sigma$
confidence level if
\begin{equation}
  (\chi^2_{\rm fixed}-\chi^2_{\rm free})/\chi^2_{\nu,{\rm free}} > 9.03,
\end{equation}
or, equivalently,
\begin{equation}
  \chi^2_{\rm fixed}/\chi^2_{\rm free} = F_\chi/dof + 1 > 1.0063,
\end{equation}
where $dof$ is the number of degrees of freedom when \vfe\ is free to vary
(1435, in this case).%
\footnote{
Note that the ratio of \ks\ adopted in \citet{sulentic12}, 1.24, is much
larger than the value derived here.  This led those authors to conclude that
the best-fit value of 730 \kms\ for the velocity shift of their B1 subsample
is not distinguishable from zero shift. It is not clear to us how they derive
this large value for the \ks\ ratio.
}

We fix \vfe\ from $-$200 to 2000 \kms, obtain the best-fit chi-square
$\chi^2_{\rm fixed}$ for each fixed input value of \vfe, and build up a \ks\
curve for each composite spectrum, as shown in the right column of Figure
\ref{fig-feschi} (crosses). For convenience, the chi-squares are plotted as
$(\chi^2_{\rm fixed}-\chi^2_{\rm free})/\chi^2_{\nu,{\rm free}}$. The
equivalent values of $\chi^2_{\rm fixed}/\chi^2_{\rm free}$ are also labeled
on the right of the plot, for comparison with Figure 1 of \citet{sulentic12}.
The solid square marks the best-fit result when \vfe\ is free to vary. The
insert zooms in around the minimum of the \ks\ curve. The 3$\sigma$
confidence interval of the \vfe\ measurement can be determined by the two
points when $(\chi^2_{\rm fixed}-\chi^2_{\rm free})/\chi^2_{\nu,{\rm free}}$
reaches 9.03 (marked by the horizontal dashed line). Fixing \vfe\ to values
outside of this velocity interval gives worse \ks\ than setting \vfe\ free, at
3$\sigma$ confidence level. The best-fit value of \vfe\ when it is free to
vary and its lower and upper 3$\sigma$ bounds are labeled in the panel on the
left column.

For all five composite spectra the $\chi^2_{\rm fixed}$ when \vfe\ is fixed to
zero are significantly larger than $\chi^2_{\rm free}$ when \vfe\ is allowed
to vary. Thus, the \feii\ emission in the composite spectra \emph{do} exhibit 
significant velocity shift. The resultant values of \vfe\ (and its 3$\sigma$
confidence level) for the five composite spectra are $210_{-70}^{+60}$,
$550_{-100}^{+100}$, $890_{-220}^{+230}$, $1110_{-250}^{+290}$, and
$1480_{-370}^{+440}$ \kms. The measured \vfe\ of the first three composite
spectra are consistent with the velocity shift range of the bins, while those
of the last two are slightly lower. This discrepancy is probably caused by the
enhancement of the un-shifted spectral features (e.g., narrow emission lines
and host galaxy component) during the stacking. Figure 12 of \citet{hu08b}
shows that \caii\ absorption lines, produced by host galaxy starlight, are
prominent in composites D and E. This shortcoming of the composite spectra has
already been discussed in \S 4.6 of \citet{hu08b}; it also explains why the
velocity shifts of \feii\ cannot be seen by direct visual inspection.

\heii\ emission and host galaxy contamination may affect the \vfe\ measurement
in some individual cases. A thorough analysis of this problem, taking both
factors into consideration, is beyond the scope of this paper. However, the
fact that \vfe\ is seen both in the composite spectra and it has a value
similar to the velocity shift range used to stack the composite spectra
suggest that the measurement of \vfe\ in \citet{hu08b} is reliable for the
majority of quasars. The values of \vfe\ are thus suitable for the SPCA study
in this paper. In fact, \citet[][their \S 3.3]{hu08b} had already tested the
reliability of their \vfe\ measurements for the S/N of SDSS spectra.
Moreover, the influence of \heii\ emission is mitigated by the fact that
\feii\ is fitted over a rather wide wavelength range.

We do not have a definitive explanation for the contradictory results obtained
by \citet{sulentic12}. Part of the problem may lie in the manner in which they
constructed their composite spectra, which are defined by parameters of their
4D Eigenvector 1 formalism, namely \hb\ width and \feii/\hb\ intensity ratio.
Figure 9 of \citet{hu08b} shows that these parameters are poorly correlated
with \vfe. Thus, composite spectra generated in bins of these spectral
properties are not equivalent to those constructed from bins in \vfe. We
suspect that this may be the reason that \citet{sulentic12} failed to see the
\feii\ velocity shifts reported by \citet{hu08b}.

\section{Derivation of the Eigenspectra}
\label{app-derivation}

This Appendix describe the detailed algorithm for deriving the eigenspectra.
The algorithm basically follows that in \citet{yip04a}.

After the three steps of spectra preprocessing described in \S
\ref{sec-pcamath}, we prepare three $816 \times 1962$ matrices: $\mathbf{F}$,
$\boldsymbol{\Sigma}$, and $\mathbf{M}$, where $\mathbf{F}$ has element
$f_{ij}$ as the flux of the $i$th mean-subtracted spectrum in the $j$th
wavelength bin, and $\boldsymbol{\Sigma}$ and $\mathbf{M}$ specify the
corresponding flux errors and masks given by the SDSS spectra. The goal of
SPCA is to calculate the eigenvalues and eigenvectors of the correlation
matrix $\mathbf{F}^T \cdot \mathbf{F}$:
\begin{equation}
  (\mathbf{F}^T \cdot \mathbf{F}) \cdot \mathbf{V} = 
  \mathbf{V} \cdot (\boldsymbol{\Lambda}^T \cdot \boldsymbol{\Lambda}),
\end{equation}
where $\mathbf{F}^T$ denotes the transpose of matrix $\mathbf{F}$,
$\mathbf{V}$ is an $n \times n$ orthogonal matrix whose columns
($\mathbf{v}_k$, $k = 1,\ n$) are the eigenvectors of $\mathbf{F}^T \cdot
\mathbf{F}$, $\boldsymbol{\Lambda}$ is an $n \times n$ diagonal matrix
diag($\lambda_1\cdots\lambda_n$), and $\lambda_k^2$ are the eigenvalues of
$\mathbf{F}^T \cdot \mathbf{F}$. This can be achieved using singular value
decomposition (SVD):
\begin{equation}
  \label{eqn-svd}
  \mathbf{F} = \mathbf{U} \cdot \boldsymbol{\Lambda} \cdot \mathbf{V}^T,
\end{equation}
where $\mathbf{U}$ is an $m \times n$ column orthogonal matrix. Defining
$\mathbf{W} = \mathbf{U} \cdot \boldsymbol{\Lambda}$, the mean-subtracted
spectra can be reconstructed using the eigenspectra
\begin{equation}
  \label{eqn-recon}
  \mathbf{F} = \mathbf{W} \cdot \mathbf{V}^T,
\end{equation}
where $\mathbf{W}$ is an $m \times n$ matrix, $w_{ik}$ is the weight of the
$k$th eigenspectrum for the $i$th mean-subtracted spectrum.

Next, we derive the eigenspectra iteratively, taking the errors
$\boldsymbol{\Sigma}$ and masks $\mathbf{M}$ into account.

\vspace{1em}

1. The bad pixels in the mean-subtracted spectra are initially corrected by
mean interpolation, following \citet{yip04a}. The flux of a bad pixel in any
given spectrum is replaced by the mean of the fluxes of all the other spectra
in the same wavelength bin, which is zero: 
\begin{equation}
  f_{ij}^{\rm c} = \sum_{i,\ {\rm good\ pixel}} f_{ij}~~~\bigg/ 
  \sum_{i,\ {\rm good\  pixel}} = 0,
\end{equation}
for $i,j$ that $m_{ij}$ represents bad. The second equation holds because
$\mathbf{F}$ is mean-subtracted. Actually, the choice of interpolation method
for the initial bad pixel correction does not affect the resultant final
eigenspectra. But the mean interpolation adopted here makes the convergence of
eigenspectra faster \citep{yip04a} and is quite simple (by just setting the
flux to zero here).

\vspace{1em}

2. The corrected flux matrix $\mathbf{F}^{\rm c}$ is then factorized in the
form of Equation (\ref{eqn-svd}) using the algorithm for SVD in chapter 2.6 of
\citet{press92}.  Each column ($\mathbf{v}_k$) of the derived matrix
$\mathbf{V}$ is an eigenspectrum; they are arranged in the order of decreasing
eigenvalues $\lambda_k^2$.

3. The mean-subtracted spectra are reconstructed using the eigenspectra
obtained, taking the errors and masks into account. For the $i$th spectrum, we
find the set of $a_{ik}$ (the coefficient of the $k$th eigenspectra for the
$i$th spectrum) by minimizing the quantity
\begin{equation}
  \chi^2_i = \sum_{j,\ {\rm good\ pixel}} 
  \left(\frac{f_{ij}-\sum_k a_{ik} v_{jk}} {\sigma_{ij}}\right)^2.
\end{equation}
This can be solved by Gauss-Jordan elimination (see \S 2 of
\citealt{connolly99} for details of the derivation). The reconstructed
spectrum $f_{ij}^{\rm rec} = \sum_k a_{ik} v_{jk}$. In practice, only the
first several tens eigenspectra are needed for reconstructing quasar spectra
\citep{yip04b,boroson10}. We find that the first 30 are sufficient for the
case here (Figure \ref{fig-cumpov} in \S \ref{sec-pcades}).

4. The bad pixels in $\mathbf{F}$ are corrected again, using the reconstructed
spectra obtained in the last step: $f_{ij}^{\rm c} = f_{ij}^{\rm rec}$. Then
we cycle back to step 2 with the new $\mathbf{F}^{\rm c}$.

\vspace{1em}

5. The loop in steps 2--4 iterates until the eigenspectra converge. Following
\citet{yip04a}, the commonality between the eigenspectra derived in the $n$th
iteration ($\mathbf{V}^n$) and that in the $(n-1)$th iteration
($\mathbf{V}^{n-1}$) is defined as
\begin{equation}
  {\rm Tr} (\mathbf{S}^n \cdot \mathbf{S}^{n-1} \cdot \mathbf{S}^n) / d,
\end{equation}
where Tr is the trace of a matrix, $d$ is the dimension of the two sets of
eigenspectra for comparison, and $\mathbf{S}^n$ is the sum of the projection
operators $\mathbf{V}^n$,
\begin{equation}
  \mathbf{S}^n = \sum_{k=1}^d \mathbf{v}_k^n \cdot (\mathbf{v}_k^n)^T,
\end{equation}
where $\mathbf{v}_k^n$ is the $k$th eigenspectrum derived in the $n$th
iteration. This quantity is unity if the two sets of eigenspectra are
identical, and is zero if they are disjoint. It is not necessary to compare
the entire set. We compare the first 30 eigenspectra for consistency with the
reconstruction of the spectrum in step 3. After five iterations, ${\rm Tr}
(\mathbf{S}^5 \cdot \mathbf{S}^4 \cdot \mathbf{S}^5) / 30 = 0.9997$ (see
Figure 1 of \citealt{yip04a} for comparison). Thus, in this work, all
eigenspectra are obtained with five iterations.

\section{The Fractional-contribution Spectrum}
\label{app-pov}

This paper introduces a new quantity, the fractional-contribution spectrum,
which is useful for understanding eigenspectra. Here we present its definition
in detail and compare it to other quantities widely used in the literature.

The total variation of the sample is defined as the sum of the squares of the
differences between the normalized spectra and the mean spectrum:
\begin{equation}
  var^{\rm tot} = \sum_{i,\ j} f_{ij}^2.
\end{equation}
It consists of two parts: one that accounts for the noise in the original
spectra,
\begin{equation}
  var^{\rm err} = \sum_{i,\ j} \sigma_{ij}^2,
\end{equation}
and another that describes intrinsic variation, $var^{\rm int} = var^{\rm tot}
- var^{\rm err}$. It is more straightforward to express them in percentages,
such that $P^{\rm err} = var^{\rm err} / var^{\rm tot}$ and $P^{\rm int} = 1 -
P^{\rm err}$.

The proportion of the variation accounted for by the $k$th eigenspectrum can
be calculated as
\begin{equation}
  \label{eqn-pk}
  P_k = \lambda_k^2 / \sum_l \lambda_l^2,
\end{equation}
where $\lambda_k^2$ is the eigenvalue for the $k$th eigenspectrum
\citep{mittaz90}. This is the quantity used in previous SPCA studies, and it
represents the contribution of an eigenspectrum to the total variation of the
input spectra over the entire wavelength range. Thus, the first $n$th
eigenspectra can account for $\sum_{k=1}^n P_k$ percentage of the total
variation; it is called the cumulative proportion of variation and increases
monotonically with $n$. When it reaches $P^{\rm int}$, it indicates that the
first $n$ eigenspectra are sufficient for explaining the intrinsic variation
in the sample, and the remaining higher order eigenspectra contribute only to
noise and can be ignored for our purposes (see also \S \ref{sec-pcades}).

The quantities defined above were widely used in previous SPCA studies, but
they only refer to the average properties of the eigenspectra over the entire
wavelength range. In this work, we develop these quantities to investigate the
eigenspectra in each wavelength bin, as follows. 

First, the total and noise variation in the $j$th wavelength bin can be simply
defined as
\begin{equation}
  var_j^{\rm tot} = \sum_i f_{ij}^2,
  ~~~var_j^{\rm err} = \sum_i \sigma_{ij}^2;
\end{equation}
and
\begin{equation}
  p^{\rm err}_j=var_j^{\rm err}/var_j^{\rm tot},
  ~~~p^{\rm int}_j=1-p^{\rm err}_j.
\end{equation}

Second, after using the first $n$th eigenspectra for reconstruction, the
residual variation in the $j$th wavelength bin is the sum of the square of the
differences between the reconstructed spectra and mean-subtracted spectra in
the specific wavelength bin:
\begin{equation}
  var_{jn}^{\rm res} = 
  \sum_i \left(f_{ij} - \sum_{k=1}^n a_{ik} v_{jk}\right)^2.
\end{equation}
The proportion of the variation accounted for by the $n$th eigenspectrum in
the $j$th wavelength bin is
\begin{equation}
  \label{eqn-pov}
  p_{jn} = \frac{var_{j(n-1)}^{\rm res} - var_{jn}^{\rm res}}
  {var_j^{\rm tot}}.
\end{equation}
As in the case applied to the entire wavelength range, these values are
compared with $p_j^{\rm err}$ to determine the significance of each
eigenspectrum. 

\begin{figure}
  \centering
  \includegraphics[width=0.45\textwidth]{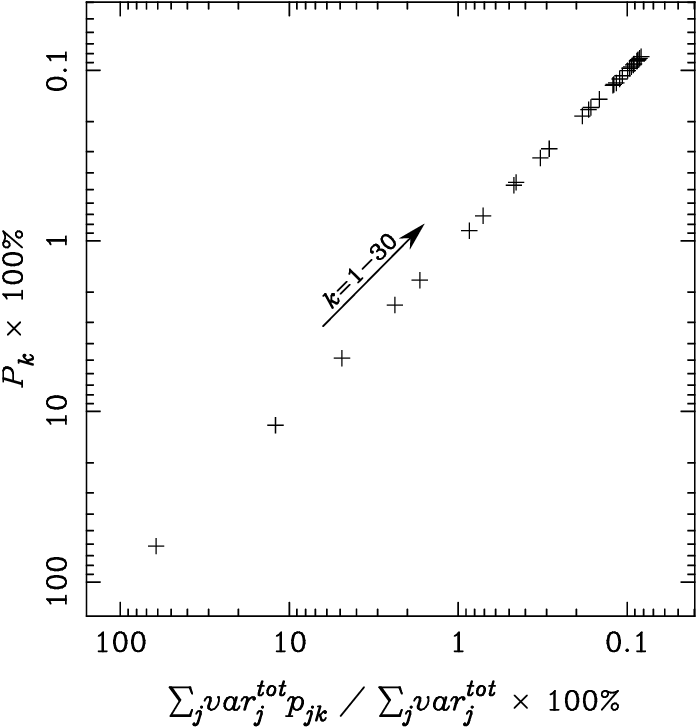}
  \caption{The sum of $p_{jk}$, weighted by $var_j^{\rm tot}$, versus $P_k$,
  expressed in percentage. $k$ increases from 1 to 30 from bottom left to top
  right. The two quantities, derived from our SPCA results, are almost equal.
  This implies that the previously used quantity $P_k$ can be considered as
  the $var_j^{\rm tot}$-weighted sum of $p_{jk}$. The new quantity $p_{jk}$
  has the advantage that it contains more diagnostic power.}
  \label{fig-pov}
\end{figure}

Thus, we obtain a matrix $\mathbf{P}$, which has the same shape as
$\mathbf{V}$, whose element $p_{jk}$ represents the proportion of the
variation accounted for by the $k$th eigenspectrum in the $j$th wavelength
bin. Each column ($\mathbf{p}_k$) of $\mathbf{P}$ represents the contribution
of the corresponding eigenspectrum and has the form of spectrum, we call it
\emph{the fractional-contribution spectrum}. Actually, as shown in Figure
\ref{fig-pov}, the previously used $P_k$ can be considered as the sum of
$p_{jk}$ weighted by $var_j^{\rm tot}$. The two quantities for the first 30
eigenspectra of our SPCA results are almost exactly equal ($k$ increases
diagonally to the upper right).

\end{document}